\documentclass[%
 reprint,
 amsmath,amssymb,
 aps,
]{revtex4-1}

\usepackage{graphicx}
\usepackage{dcolumn}
\usepackage{bm}

\usepackage{natbib}
\usepackage{graphicx}
\usepackage{epstopdf, epsfig}
\usepackage{mathrsfs}
\usepackage{graphicx,amssymb,amsmath,color}
\usepackage{soul}
\usepackage{mathtools}

\newcommand{\fig}{Fig.~}

\newcommand{\eq}{Eq.~}
\newcommand{\eqs}{Eqs.~}

\definecolor{blue2}{rgb}{0, 0.4470, 0.7410}
\definecolor{blue3}{rgb}{0.1804, 0.0784, 0.5529}
\definecolor{red2}{rgb}{0.8500, 0.1250, 0.0480} 
\definecolor{orange2}{rgb}{0.8500, 0.3250, 0.0980} 
\definecolor{yellow2}{rgb}{0.9290, 0.6940, 0.1250}
\definecolor{purple2}{rgb}{0.4940, 0.1840, 0.5560}
\definecolor{green2}{rgb}{0.4660, 0.6740, 0.1880}
\definecolor{ltblue2}{rgb}{0.3010, 0.7450, 0.9330}
\definecolor{dkred2}{rgb}{0.6350, 0.0780, 0.1840}
\definecolor{gray2}{rgb}{0.22, 0.22, 0.3}
\definecolor{gray3}{rgb}{0.5, 0.5, 0.5}

\definecolor{gray}{rgb}{0.7,0.7,0.7}

\begin{document}

\preprint{APS/123-QED}

\title{Network community-based model reduction for vortical flows}

\author{Muralikrishnan Gopalakrishnan Meena}
 \email{mg15h@my.fsu.edu}
\author{Aditya G.~Nair}
 \email{agn13@my.fsu.edu}
\author{Kunihiko Taira}
 \email{ktaira@fsu.edu}
\affiliation{%
 Department of Mechanical Engineering, Florida State University, Tallahassee, Florida, 32310, USA 
}%


\date{\today}

\begin{abstract}
A network community-based reduced-order model is developed to capture key interactions amongst coherent structures in high-dimensional unsteady vortical flows.  The present approach is data-inspired and founded on network-theoretic techniques to identify important vortical communities that are comprised of vortical elements that share similar dynamical behavior.  The overall interaction-based physics of the high-dimensional flow field is distilled into the vortical community centroids, considerably reducing the system dimension. Taking advantage of these vortical interactions, the proposed methodology is applied to formulate reduced-order models for the inter-community dynamics of vortical flows, and predict lift and drag forces on bodies in wake flows. We demonstrate the capabilities of these models by accurately capturing the macroscopic dynamics of a collection of discrete point vortices, and the complex unsteady aerodynamic forces on a circular cylinder and an airfoil with a Gurney flap.  The present formulation is found to be robust against simulated experimental noise and turbulence due to its integrating nature of the system reduction.
\end{abstract}

\pacs{Valid PACS appear here}
\maketitle


\section{Introduction}

The field of network science has paved the way for analyzing complex interactions amongst a set of connected elements \cite{Bollobas98, Newman:SIAMReview03, BarabasiNS16}.  Theoretical and computational toolsets developed in this field have inspired in-depth studies on the structural and dynamical characteristics of various networks, including biological, social, transportation, information, and power-grid networks \cite{zhu2007getting, sporns2011human, otte2002social, albert2004structural}.  In general, these networks have well-defined graph structures comprised of sets of discrete nodes.  With such graph structure at hand, graph theory, linear algebra, data science, dynamical systems theory, and control theory can be applied in harmony to study complex networked dynamics \cite{boccaletti2006complex,newman2011structure}. For example, one can examine problems such as disease and information transmission on social networks \cite{Salathe:PLOSCB10}, and development of cyber security strategies for combating attacks on the world wide web, while maintaining key connectivity \cite{Albert:Nature00,dorogovtsev2013evolution}.

Complex interactions prevalent in fluid flows also exhibit great richness in their dynamics, as seen in unsteady vortical and turbulent flows. Nonetheless, the application of network-theoretic techniques on the analysis of fluid flows remains fairly limited. One of the major challenges in extending network analysis to fluid flows is the extraction of the underlying network structure that can appropriately describe their high-dimensional kinematics and dynamics. In recent years, there have been emerging efforts to describe fluid flows with Lagragian \cite{Nair:JFM15}, Eulerian \cite{Taira:JFM16, scarsoglio2016complex}, and modal \cite{nair2017networked} network representations.  In the Lagrangian fluid flow network, the vortices can be taken to be nodes and the induced velocity magnitudes from one vortex to others can be considered as edge weights.  Such formulation allows for the preservation of conservative variables and has led to the development of a sparse vortical dynamics model \cite{Nair:JFM15}. The Eulerian network description has been applied to characterize two-dimensional turbulence in an analogous manner and has revealed its scale-free network characteristics \cite{Taira:JFM16}. Moreover, kinetic energy transfer in unsteady fluid flows has been captured with a modal network, which has served as a basis to perform feedback stabilization for drag reduction \cite{nair2017networked}.

The characterization of interactions among the vortical elements is challenging due to the high-dimensional nature of system.  Thus, there is a need to reduce the dimension of the system, while preserving the overall interaction structure and dynamics. In the present work, we reduce the order of a complex nonlinear system, particularly of unsteady wake flows, by identifying communities in the fluid flow network representing interactions among the vortical elements. Furthermore, we model their dynamical characteristics and other system outputs in a computationally tractable manner.  While several works in literature take advantage of the community structure to describe system behavior like evolving networks and mobility models \cite{xie2007new,musolesi2006community}, studies focusing on a community-based reduced-order formulation to describe the dynamics and other outputs of a system are limited in literature and the fluid mechanics community.

Community detection of networks has received attention in various data-rich fields \cite{mucha2010community, newman2006modularity, Fortunato:PR10}. Aggregating nodes of a network with high density of intra-group interactions into communities has a lot of implications in reduction of system complexity. Each community or cluster can then be utilized to describe characteristics of the global system \cite{variano2004networks}.  In the field of fluid mechanics, the connections between clusters of Lagrangian trajectories and coherent vortical structures were explored in the work of \citet{hadjighasem2016spectral}. Spectral graph drawing has also been examined to identify coherent structures \cite{schlueter2017coherent}. The methodology presented in the current work compliments these efforts in extracting communities in fluid flows purely from the underlying vortical connectivity structure, instead of fluid trajectories, with an added attribute of developing community-based reduced-order models. These communities are key flow features that are important in capturing the interactions that govern the wake dynamics. Based on the identified communities, we have developed reduced-order models to predict cluster trajectories and aerodynamic forces on bluff bodies. These are demonstrated on well-know vortical flow problems. The use of network-based concepts to describe the complex nature of fluid flows is attractive, especially since they have the ability to capture the interactive dynamics and lead to models that are amicable to preserving key interactions in a sparse manner.

In what follows, we first discuss the general approach for the community-based dimension reduction of networked systems in Sec.~\ref{sec:approach}. We describe the procedure for modeling the overall networked dynamics of the system and predicting observable (auxiliary) variables of the system. The formulations are then extended to analyze unsteady vortical flows, which is elaborated in Sec.~\ref{sec:ext_fluid}.  The vortical flows are mathematically cast in terms of vortical networks and are dimensionally reduced by identifying community structures within these networks. With a reduced community-based representation of the vortical flows, we develop a nonlinear reduced-order model to predict the overall dynamics of the communities and also the aerodynamic forces, which are described in Sec.~\ref{sec:application}. We demonstrate the capability of the present community-based reduced-order model using canonical two-dimensional vortical flows and discuss its robustness against perturbations in the flow field data. While the present paper discusses the formulation in the context of vortical flows, the overall approach is applicable to a wide range of networked dynamics and continuum mechanics problems.


\section{Approach}
\label{sec:approach}

Let us consider a dynamical system on a network (graph $\mathcal{G}$) described by an adjacency matrix $A \in \mathbb{R}^{n\times n}$.  On each node of the network, we have a state vector $\boldsymbol{x}_i \in \mathbb{R}^p$ holding $p$ variables being analyzed.  We consider this state variable to evolve through the networked dynamics described by 
\begin{equation}
   \dot{\boldsymbol{x}}_i = \boldsymbol{f}(\boldsymbol{x}_i) + \sum_{j=1}^n A_{ij} \boldsymbol{g}(\boldsymbol{x}_i, \boldsymbol{x}_j),
   \quad i = 1, 2, \dots, n 
   \label{eq:1}
\end{equation}
where $\boldsymbol{f}$ and $\boldsymbol{g}$ are in general nonlinear functions \citep{Newman:10}.  Function $\boldsymbol{f}(\boldsymbol{x}_i)$ represents the intrinsic dynamics at node $i$ and function $\boldsymbol{g}(\boldsymbol{x}_i,\boldsymbol{x}_j)$ describes the interactive dynamics that takes place between nodes $i$ and $j$.  The adjacency matrix $A_{ij}$ in the above equation indicates the existence of the interactions with unweighted or weighted edges between nodes $i$ and $j$.  In many cases, the adjacency matrix $A$ is known from physics or the governing equation. 
For systems with no {\it a priori} knowledge of the network structure, one can find the adjacency matrix $A$ based on system identification or network inference \cite{kramer2009network, hecker2009gene, de2010advantages, nelles2013nonlinear}.

For many problems, networks are comprised of a large number of nodes (i.e., $n \gg 1$).  The analysis, simulation, and modeling of dynamics on such large networks requires significant computational resources and calls for techniques to reduce the computational complexity of the problem, while retaining the essence of the key physics.  In the present approach, we classify the nodes into network communities based on the adjacency matrix $A$ and reduce the state dynamics of the overall system to the community-based (centroid-based) dynamics.  In the discussion herein, a network community is considered as a group of nodes that are strongly connected and behave collectively in a dynamically similar manner.  

\begin{figure*}
  \centerline{\includegraphics[width=0.9\textwidth]{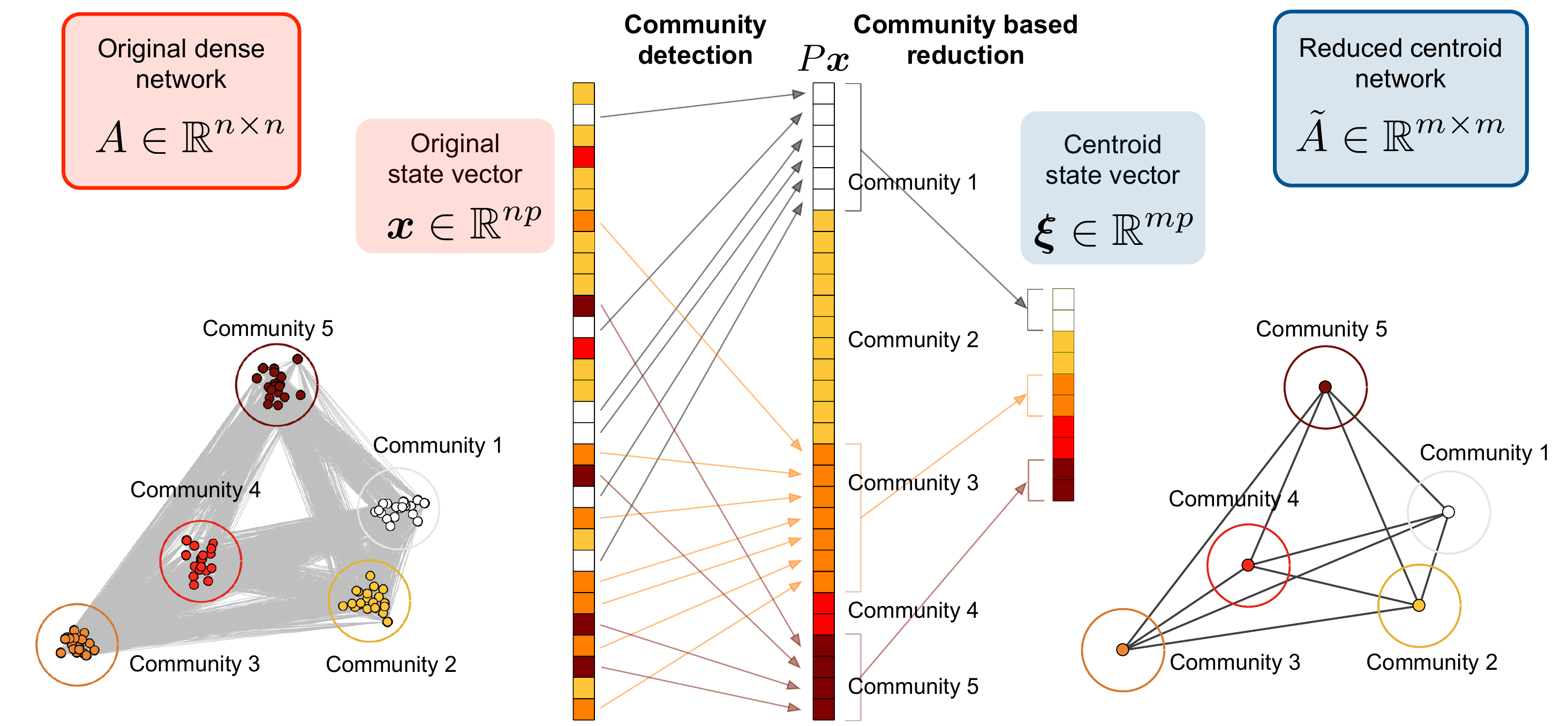}}
  \caption{Overview of the community-based reduction of a networked dynamical system.}
  \label{fig:overview}
\end{figure*}

We illustrate the overall procedure of the present community-based model reduction technique in \fig \ref{fig:overview}.  Given the adjacency matrix $A$, we perform community detection.  There are a number of community detection algorithms which can be utilized.  One of the most popular community detection techniques is the modularity maximization algorithm \citep{newman2004finding, newman2004fast, newman2006modularity}. Nodes in a particular community of a network interact in a similar manner compared to the rest of the nodes or communities in the network. Modularity is a measure representing the difference between the fraction of edges within a community and the expected value for a randomly generated network with same size and degree distribution. Modularity, $Q$ for a directed network is defined as,
\begin{equation}
   Q = \frac{1}{2n_e}\sum\limits_{ij}
   \left[ A_{ij} - \frac{s_i^\text{in}s_j^\text{out}}{2 n_e} \right]\delta(c_i,c_j),
\end{equation}
where $s^{\text{in}}_i = \sum_{j=1}^n A_{ij}$ and $s_j^\text{out} = \sum_{i=1}^n A_{ij}$ are the in and out-degrees (strengths), respectively, $n_e$ is the total number of edges in the network, $\delta(i,j)$ is the Kronecker delta, $c_i \in C_l$ is the label of the community to which element $i$ is assigned and $C_l$ is the set of $l$-th network community \citep{leicht2008community}. Here, $l = 1,2, \dots, m$, with $m$ being the total number of communities. Higher positive values of modularity indicate more edges within communities than what is expected on the basis of chance, portraying a high modular structure of the network.  Thus, the algorithm arranges the elements of the network into communities so as to maximize the modularity measure of the network. This approach allows for communities to be identified without specifying their sizes or their number within the overall network. Note that community detection defers from graph partitioning \cite{Newman:10}.  While modularity maximization is only one of the techniques to perform community detection, other techniques such as using eigenvectors of matrices, the Louvain method, spectral methods, and using random walks \cite{newman2006finding,blondel2008fast,newman2013spectral,pons2006computing,zhang2015multiway} can be applied to detect community structure depending on the problem at hand.

Now, let us consider an aggregate state vector $\boldsymbol{x}$ comprised of the state vectors from all nodes $\{ \boldsymbol{x}_i \}_{i=1}^n$:
\begin{equation}
   \boldsymbol{x} \equiv 
   \begin{pmatrix} \boldsymbol{x}_1 \\ \boldsymbol{x}_2 \\ \vdots \\ \boldsymbol{x}_n \end{pmatrix}
   \in \mathbb{R}^{np}.
\end{equation}
With the network communities identified, we can order the above aggregate state variable to sort the entries based on the communities to which they belong.  It is possible to construct a permutation matrix $P$ equivalently that would take the left vector $\boldsymbol{x}$ to the middle vector $P \boldsymbol{x}$ in \fig \ref{fig:overview}.  While this reordering is performed here for visual clarity, it is not necessary in programming implementation, as long as the indices are tracked.  The main objective here is to determine the reduced graph $\tilde{\mathcal{G}}$ representing the interactions amongst the communities. 

With the elements of $\boldsymbol{x}$ grouped into respective communities, we can reduce the state variables of each communities to their respective centroids.  That is, we define a centroid of community $k$ based on an appropriate weighing variable of the system, $\kappa_i$, 
\begin{equation}
   \boldsymbol{\xi}_k \equiv 
   \frac{ \sum_{i \in C_k} \kappa_i \boldsymbol{x}_i}{ \sum_{i \in C_k} \kappa_i },
   \quad k = 1, 2, \dots, m,
   \label{eq:comm_avg}
\end{equation}
where $m \ll n$.
Accordingly, we can define an aggregate community centroid-based state vector $\boldsymbol{\xi}$ 
\begin{equation} 
   \boldsymbol{\xi} \equiv 
   \begin{pmatrix} \boldsymbol{\xi}_1 \\ \boldsymbol{\xi}_2 \\ \vdots \\ \boldsymbol{\xi}_m \end{pmatrix}
   \in \mathbb{R}^{mp},
\end{equation}
which is reduced from $P\boldsymbol{x}$ in \fig \ref{fig:overview}.
We herein call this reduction {\it community-based reduction}, as we now intend to capture the overall dynamics of the system by tracking the community centroids in a macroscopic manner, instead of being concerned of each and every member of the community.

Let us now perform a community-based average (\ref{eq:comm_avg}) of the governing equation (\ref{eq:1}) to find
\begin{equation}
   \dot{\boldsymbol{\xi}}_k 
   = \frac{ \sum_{i \in C_k} \kappa_i \boldsymbol{f}(\boldsymbol{x}_i)}{ \sum_{i \in C_k} \kappa_i }+ \sum_{j=1}^n \frac{ \sum_{i \in C_k} \kappa_i A_{ij} \boldsymbol{g}(\boldsymbol{x}_i, \boldsymbol{x}_j)}{ \sum_{i \in C_k} \kappa_i }
   \label{eq:reduced}
\end{equation}
for $k = 1, \dots, m$.  It is further desirable to have the right hand side of the above equation be dependent on $\boldsymbol{\xi}$ instead of $\boldsymbol{x}$.  The reduction can further be simplified if $\boldsymbol{f}$ is a linear function, in which case, we have
\begin{equation}
\begin{split}
   \dot{\boldsymbol{\xi}}_k 
   & = \boldsymbol{f}(\boldsymbol{\xi}_k)
   + \sum_{j=1}^n \frac{ \sum_{i \in C_k} \kappa_i A_{ij} \boldsymbol{g}(\boldsymbol{x}_i, \boldsymbol{x}_j)}{ \sum_{i \in C_k} \kappa_i }\\
  & = \boldsymbol{f}(\boldsymbol{\xi}_k)
   + \sum_{l=1}^m \frac{\sum_{j \in C_l} \sum_{i \in C_k} \kappa_i A_{ij} \boldsymbol{g}(\boldsymbol{x}_i, \boldsymbol{x}_j)}{ \sum_{i \in C_k} \kappa_i }.
   \end{split}
   \label{eq:reduced2}
\end{equation}
We note that the averaging of the nonlinear term $\boldsymbol{g}$ in general cannot be further simplified.  However, we can approximate this nonlinear term as a function of $\boldsymbol{\xi}$.  That is, we can reduce the governing equation to the form of
\begin{equation}
   \dot{\boldsymbol{\xi}}_k 
   = \boldsymbol{f}(\boldsymbol{\xi}_k)
   + \sum_{l=1}^m \tilde{A}_{kl} \tilde{\boldsymbol{g}}(\boldsymbol{\xi}_k, \boldsymbol{\xi}_l)
   \label{eq:reduced3}
\end{equation}
based on the centroid-based state variable $\boldsymbol{\xi}$.

The governing equation of the above form can be derived using two approaches.  If the networked dynamics is well known from theory, we can use the same interaction function $\boldsymbol{g}$ for $\tilde{\boldsymbol{g}}$ and the same form of the network edge weights. This implicitly assumes that the edge weight can be quantified in the same manner. It should be noted that reduced adjacency matrix $\tilde{A} \in \mathbb{R}^{m\times m}$ is much smaller in size compared to the original matrix $A \in \mathbb{R}^{n\times n}$.  While this approach appears simple, the use of the original forms for $\tilde{\boldsymbol{g}}$ and $\tilde{A}$ generally leads to an approximate model.  

An alternative approach is to determine $\tilde{A}$ and $\tilde{\boldsymbol{g}}$ numerically through system identification \cite{nelles2013nonlinear}.  Data-based techniques can be used to find the reduced adjacency matrix and the interaction function if time-series data of the dynamical systems are available for $\boldsymbol{\xi}(t)$.  One can pursue algorithms that minimizes the $L_2$ measure or the $L_1$ norm to promote sparsity within the derived model.  Nonlinear system identification method, such as SINDy (sparse identification of nonlinear dynamics), has enabled the sparse identification of governing equations \cite{Brunton:PNAS16}.  In the examples we consider later, we use both of the aforementioned approaches.

By establishing the community-based model of the networked dynamics, we can extend the present approach to model an observable or auxiliary variable $\boldsymbol{y} \in \mathbb{R}^q$ that is dependent on the state dynamics.  
That is, for such variables that satisfies
\begin{equation}
   \boldsymbol{y} = \boldsymbol{h}(\boldsymbol{x})
   \quad \text{or} \quad
   \dot{\boldsymbol{y}} = \boldsymbol{h}(\boldsymbol{x},\boldsymbol{y}),
\end{equation}   
we seek a reduced-order model 
\begin{equation}
   \boldsymbol{y} = \tilde{\boldsymbol{h}}(\boldsymbol{\xi}),
   \quad \text{or} \quad
   \dot{\boldsymbol{y}} = \tilde{\boldsymbol{h}}(\boldsymbol{\xi},\boldsymbol{y})
   \label{eq:y_reduced}
\end{equation}   
based on the centroid variable $\boldsymbol{\xi}$.  The function $\tilde{\boldsymbol{h}}$ can be modeled through nonlinear regression analysis \citep{bates1988nonlinear}.  Again, it is possible to develop a model based on the use of different norms as mentioned above.  To fit an appropriate model, a library of the possible nonlinear basis functions $\boldsymbol{\Theta}(\boldsymbol{\xi}, \boldsymbol{y})$ can be constructed so that
\begin{equation}
\dot{\boldsymbol{y}} = \boldsymbol{\Theta}(\boldsymbol{\xi}, \boldsymbol{y})\boldsymbol{B}.
\label{eq:y_regression}
\end{equation}
For example, $\boldsymbol{\Theta}$ can consist of a library of polynomials and trigonometric terms of $\boldsymbol{\xi}$ and $\boldsymbol{y}$ \cite{nelles2013nonlinear,Brunton:PNAS16}. These terms, along with $\dot{\boldsymbol{y}}$, are compiled from a time series of data for a training period of $-T \le t \le 0$. From such a library, the regression coefficients $\boldsymbol{B}$ can be determined by a matrix solver. Moreover, it is possible to consider sparse regression for solving $\boldsymbol{B}$ following the sequential thresholding least-squares algorithm implemented in SINDy \cite{Brunton:PNAS16}, further simplifying the governing equation in the case of $\dot{\boldsymbol{y}} = \tilde{\boldsymbol{h}}(\boldsymbol{\xi},\boldsymbol{y})$.  Once these models are developed, they can be used to estimate the output variable $\boldsymbol{y}$ or predict it based on training data and a given initial condition.  

The implications of the formulations discussed above pertains towards analyzing the overall dynamics of high-dimensional systems with an intrinsic community structure. Also, this is achieved solely through utilizing data inspired network-based techniques taking advantage of the interaction-based physics.  With such techniques, one need not track all elements involved in the system, whereas just capture the community centroids in the system to uncover the overall networked dynamics and other auxiliary variables of the system. Below, we demonstrate the effectiveness of the present formulation to capture the macroscopic motion of vortices in nonlinear flows. Two examples are considered to highlight the ability to predict vortex community motion and to develop models that can capture aerodynamic forces on bodies imposed by the unsteady wake flows based on extracted vortical communities.

\section{Extension to Fluid Flows}\label{sec:ext_fluid}
\subsection{Network representation of the Biot--Savart law}\label{subsec:biot_net}

Fluid flows pose one of the most nonlinear, multi-scale phenomena in nature. The Navier--Stokes equations that describe the high-dimensional dynamics of incompressible flows are
\begin{align}
   \boldsymbol{u}_t
   + \boldsymbol{u}\cdot\nabla\boldsymbol{u} 
   &= -\nabla p + Re^{-1} \nabla^2\boldsymbol{u},\\
   \nabla \cdot \boldsymbol{u} &= 0, 
\end{align}
where $\boldsymbol{u}$ represents the velocity, $p$ denotes the pressure, and $Re$ is the Reynolds number. The above formulation can be equivalently represented for the vorticity field $\boldsymbol{\omega} = \nabla \times \boldsymbol{u}$ with
\begin{equation}
   \boldsymbol{\omega}_t 
   + \boldsymbol{u}\cdot \nabla \boldsymbol{\omega} 
   = \boldsymbol{\omega} \cdot \nabla \boldsymbol{u} 
   + Re^{-1} \nabla^2 \boldsymbol{\omega}.
\end{equation}
As the vorticity field describes unsteady motion of fluid flows, we take the viewpoint that key flow features are represented as vortical elements in the present study.

These equations governing the dynamics of the vortical elements in a flow are complex to solve due to the presence of the nonlinear terms, requiring a full computational fluid dynamics solver \cite{Kajishima17}. Resolving the full dynamics through numerical as well as experimental methods requires resources and sometimes pose major challenges, especially when the Reynolds number is high, giving rise to turbulence. For this reason, we utilize the network community-based reduced-order formulations, taking advantage of the interaction-based physics, to model the dynamics and other auxiliary variables based on the inter-community dynamics.

The vortical elements influence each other by imposing induced velocity onto each other. The interaction strength depends on both the distance and circulation of the vortex elements \cite{Saffman:92}. We capture these interactions in fluid flows through a vortical network $\mathcal{G}$ where the nodes represent the vortex elements and the edges with weights $\mathcal{W}$ represent the strength of interactions between the vortex elements. 

\begin{figure}
  \centerline{\includegraphics[width=0.48\textwidth]{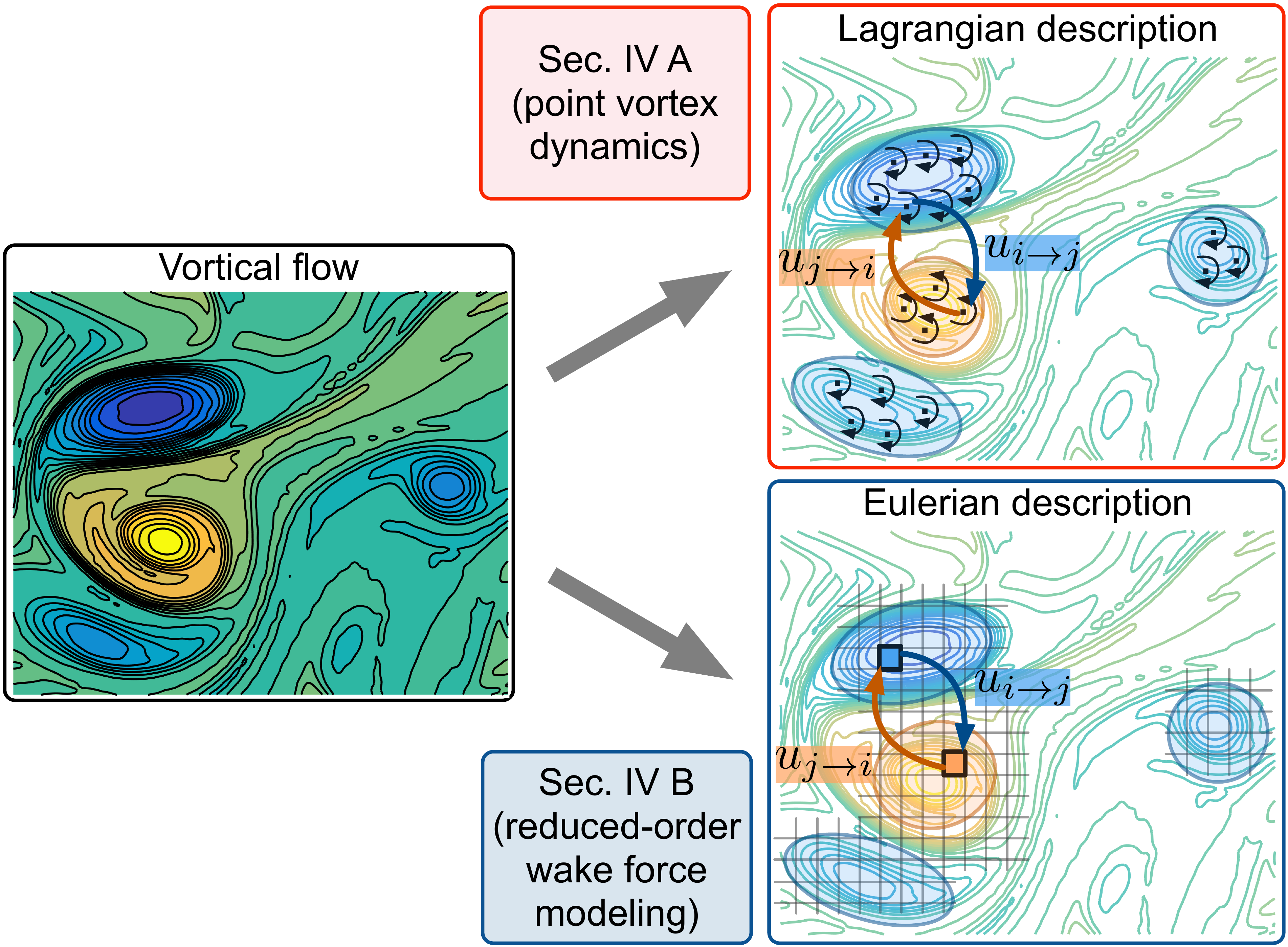}\quad}
  \vspace{-0.05in}
  \caption{Lagrangian and Eulerian network descriptions of a flow field. Nodes (discrete point vortices and Cartesian cells) and edges (induced velocity $u$) of the vortical network in each descriptions are also shown.}
\label{f:flow_rep}
\end{figure}

Vortical elements on the network can be described through a Lagrangian or Eulerian perspective as shown in \fig \ref{f:flow_rep}. In the Lagrangian description, the vortical elements are approximated using a collection of discrete point vortices (line vortices), which serve as nodes of the vortical network.  The motion of the vortical structures can be captured by modeling the dynamics of these point vortices. The number of nodes depend on the number of point vortices used to discretize the vortical structures in the flow field. The strength of each point vortex is given by the circulation corresponding to the local area covered by the point vortex while discretizing the vortical structure.

A continuous representation of the flow field can also adopt an Eulerian description, where Cartesian discretization of the flow with vortical elements within the grids correspond to nodes of the network, as shown in \fig \ref{f:flow_rep}. The vortical elements within each grid cell keep changing with time according to the flow dynamics. The circulation over the grid cell is used as the circulation of the node. To highlight the generality of our current approach, we include discussions on both network formulations in Sections \ref{subsec:point_vortex_dyn} and \ref{subsec:force_model}.

The influence of vortical elements governed through their induced velocity makes it a natural choice to define the strength of interaction among the vortical nodes over the fluid flow network. The induced velocity can be determined by the Biot--Savart's law \cite{Saffman:92}, which for an element $i$ in a two-dimensional, inviscid, incompressible flow field is given by
\begin{equation}
\boldsymbol{u}(\boldsymbol{r}_i,t) =
\frac{{\rm d}{\boldsymbol{r}_i}}{{\rm d}t} =
\sum\limits_{\substack{j=1\\j \ne i}}^n \frac{\gamma_j}{2 \pi} \frac{\hat{\boldsymbol{k}} \times (\boldsymbol{r}_i - \boldsymbol{r}_j)}{|\boldsymbol{r}_i - \boldsymbol{r}_j|^2},
\label{eq:BS}
\end{equation}
where $\boldsymbol{u}(\boldsymbol{r}_i,t)$ is the induced velocity on element $i$ by rest of the $n$ vortical elements in the flow field, $\boldsymbol{r}$ and $\gamma$ are the position vector and circulation of the vortex elements, respectively, and $\hat{\boldsymbol{k}} $ is the out-of-plane unit normal vector. Here, a vortical element $i$ cannot induce velocity upon itself. The interactions among vortical elements for both Lagrangian and Eulerian descriptions are captured by quantifying the induced velocity.  The Biot--Savart law is used to quantify the vortical interactions to construct a model of the dynamics of the fluid flow governed by the Navier--Stokes equations. Although, \eq (\ref{eq:BS}) can be used to study the dynamics of vortical elements, as described later in Sec.~\ref{subsec:point_vortex_dyn}.

Based on the above relationship, the induced velocity from vortex element $i$ onto element $j$ can be simplified to
\begin{equation}
{u}_{i\rightarrow j} =
 \frac{| \gamma_i |}{2\pi |\boldsymbol{r}_j - \boldsymbol{r}_i|},
 \quad
 i \ne j,
 \label{eq:BS_simple}
\end{equation}
as depicted in \fig \ref{f:flow_rep}. Note that the absolute value of $\gamma$ is used as the network-based techniques, like community detection, are limited for signed graphs \cite{traag2009community, esmailian2015community,sugihara2013community}. Although, this can be relaxed during the definition of the vortical network. We separate the data with opposite signed circulation while performing such techniques to account for the differences in vortical behavior. This is further discussed in the Sec.~\ref{subsec:fluid_community}.

In our formulation, state vector $\boldsymbol{x}$ for the generalized dynamical system in \eq (\ref{eq:1}) holds the position vector $\boldsymbol{r}$ for fluid flow application. Moreover, \eq (\ref{eq:BS}) can be related back to the networked dynamics equation, \eq (\ref{eq:1}), with $\boldsymbol{f}(\boldsymbol{r}_i) = 0$ and
\begin{equation}\boldsymbol{g}(\boldsymbol{r}_i, \boldsymbol{r}_j) = \frac{\hat{\boldsymbol{k}} \times (\boldsymbol{r}_i- \boldsymbol{r}_j)}{|\boldsymbol{r}_i- \boldsymbol{r}_j|}.
\label{eq:gorg}
\end{equation}
The weighted adjacency matrix $A \in \mathbb{R}^{n\times n}$ for this vortical flow network \cite{Nair:JFM15,Taira:JFM16} is defined as 
\begin{equation}
  A_{ij} = \left\{
    \begin{array}{ll}
      w_{ij}   & \text{if~} i \neq j \\
      0        & \text{if~} i = j,
  \end{array} \right.
  \label{eq:adj_org}
\end{equation}
where $w_{ij}$ is the edge weight given by 
\begin{equation}
w_{ij} = \alpha u_{i\rightarrow j} + (1-\alpha) u_{j\rightarrow i}
\end{equation}
with $\alpha \in [0,1]$ being a parameter to capture the induced velocity direction to construct $\mathcal{W}$.  With $\alpha = 0$ or $1$, we have a directed adjacency matrix. One can also select $\alpha = 1/2$ to yield a symmetric network \cite{Taira:JFM16}.  In the current work, we set $\alpha = 0$.

In the current method, a snapshot of the velocity or vorticity field of any two-dimensional, incompressible fluid flow problem is solely required to represent the flow field in the above network formulation, as shown in \fig \ref{f:flow_rep}. We have observed that the inviscid approximation of the Biot--Savart law is able to capture the interaction physics in viscous flows effectively through this network definition \cite{Nair:JFM15,Taira:JFM16}. Also, the network representations of time series of snapshots can describe the time evolution of interactions among the vortical structures. To reduce the number of elements analyzed, we concentrate only on the core vortical structures or coherent structures in a flow field, isolated through vorticity thresholding. Nonetheless, the number of elements in such a system, dependent on the grid resolution of the flow field, can still be high, making the dynamics tracking a computationally challenging problem. We apply the community-based reduction procedure on the network representation of such high-dimensional fluid flow system to reduce its order.

\subsection{Community-based reduction of vortical flows}\label{subsec:fluid_community}

To reduce the dimension of the system, we distill the vortex communities (coherent network structures) to their controids. We use the community detection methodology described in Section \ref{sec:approach} to identify these coherent structures or vortical communities from the given data. The segregation of vortical elements behaving or interacting in similar manner to form a vortex structure or cluster can be defined as a vortical community from a network-theoretic perspective. In this study, the definition of a vortical community need not necessarily align with the definition of a vortex in the fluid dynamics literature, since the present objective is to construct a network based model to reduce the order of the system and develop a network-based reduced dynamical model.

The procedure for identifying communities in the vortical network of a collection of discrete point vortices is portrayed in \fig \ref{fig:overview}. For a given snapshot of any two-dimensional flow field, we first perform vorticity thresholding to remove non-vortical region of the flow and decompose the flow field into positive and negative vorticity fields.  Next, a directed version of the modularity maximization algorithm \cite{newman2004analysis,clauset2004finding,leicht2008community} is applied to the two data sets comprising of vortical elements with clockwise (negative) and counter-clockwise (positive) vorticity. Due to the unsigned edge weight, we need to separate the two data sets as the current community detection procedure is unable to distinguish spatially close vortical communities, usually the positive and negative vortical structures, especially for wake flows. The resolution effects in the modularity maximization algorithm \cite{fortunato2007resolution,reichardt2004detecting,reichardt2006statistical} are also taken into consideration.

Once the communities are detected from the flow field, the centroid of each identified community is calculated by
\begin{equation}
   \boldsymbol{\xi}_k \equiv 
   \frac{ \sum_{i \in C_k} \gamma_i \boldsymbol{r}_i}{ \sum_{i \in C_k} \gamma_i },
   \quad k = 1, 2, \dots, m
   \label{eq:vort_centroid}
\end{equation}
and the total circulation of each community 
\begin{equation}
   \Gamma_k = \sum_{i \in C_k} \gamma_i
\end{equation}
is concentrated to the community centroid, distilling the vortical community into a point vortex at the community centroid. The vortical network is redefined using the circulation and position vector of each community centroid, thus reducing the dimension of the system significantly. Although it is not performed here, it is possible to consider a bipartite graph to relate the reduced centroid-based network to the original dense vortical network using an incidence matrix \cite{Newman:10}. 

With the formulation established for vortical flows, we can model the overall dynamics of the original high-dimensional networked fluid flow system using the present community-based approach. Furthermore, we can formulate other auxiliary variables in fluid flow problems, like unsteady lift and drag forces on bodies, solely using the properties of the community centroid variable $\boldsymbol{\xi}$.  These applications are presented in the following sections.

\section{Applications}\label{sec:application}

In this section, we take the network community-based reduced-order modeling technique described above to study two types of vortical flows.  First, we consider a group of discrete point vortices and model their collective motion based on the community centroids.  Second, we consider unsteady wake flows behind a circular cylinder and an airfoil with a flap, whose wakes are described by the motion of vortical communities.  The properties of these communities are then used to derive reduced-order models to predict the forces imposed on the bodies.  These flows, familiar in the field of fluid dynamics, are chosen to validate the present approach. These model problems highlight the capabilities of the network community-based modeling technique to significantly reduce the dimensions of the complex nonlinear flow physics, while retaining the important interactions present in the flows.

\subsection{Point vortex dynamics}\label{subsec:point_vortex_dyn}

\begin{figure*}
   \begin{center}
   \includegraphics[width=0.95\textwidth]{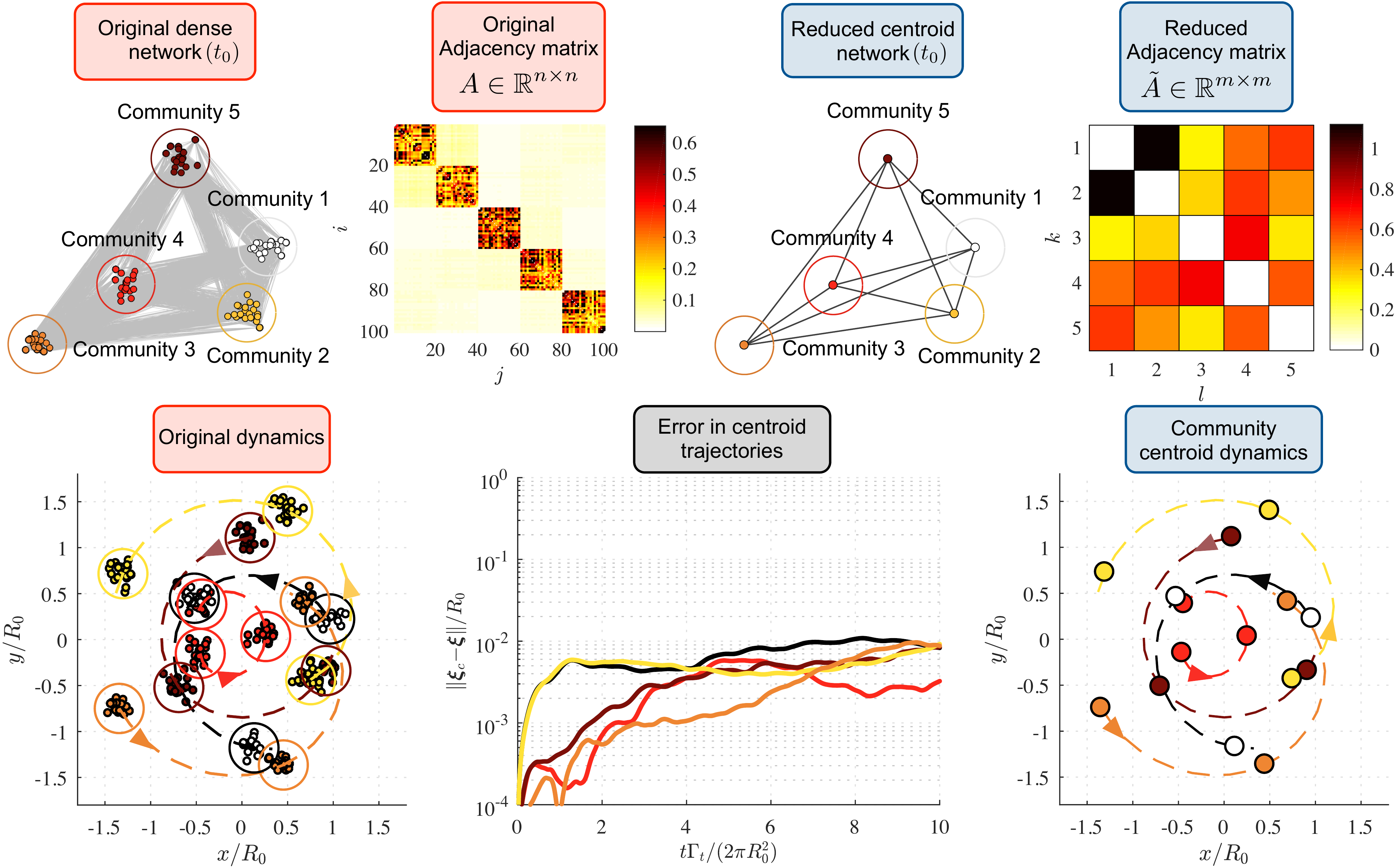}
   \end{center}
   \vspace{-0.2in}
   \caption{Community-based networked dynamics of point vortices. The network structure and adjacency matrix at initial time ($t_0$), overall dynamics, and error between the community centroid dynamics for both the original dense ($n = 100$) and the community-based reduced ($m = 5$) point vortex systems are shown.}
\label{f:pvd}
\end{figure*} 

We first study the unsteady dynamics of a collection of discrete point vortices as shown in \fig \ref{f:pvd} (also in \fig \ref{fig:overview}). As we aim to capture the time-varying behavior of a collection of vortices, the Lagrangian description is apt to construct the community-based reduced order model.  Here, we consider $n=100$ discrete point vortices and their collective motion.  These vortices are provided with circulation $\gamma_i$ having a normal distribution with a mean of $\bar{\gamma} = 0.1$ and a standard deviation of $\sigma_\gamma = 0.008$. The setup portrays the strength distribution of vortical structures closely found in many vortical flows. Based on the given group of vortices, we construct the vortex network and the adjacency matrix, where the nodes correspond to the $100$ discrete point vortices in the spacial domain and the edges are given by \eq (\ref{eq:adj_org}). This forms a complete graph (dense network) of vortical interactions as shown in \fig \ref{f:pvd} (top left) resulting from velocity induced by the vortices on each other. The vortical network does not contain any self-loops as a vortex does not induce velocity upon itself. Thus, the total number of interactions in the original vortical network is $n(n-1)$. The adjacency matrix corresponding to the vortical interactions is shown adjacent to the dense network in \fig \ref{f:pvd}. The colors in the adjacency matrix denote the edge weights $w_{ij}$.

Using modularity maximization algorithm described in Sections \ref{sec:approach} and \ref{subsec:fluid_community}, the set of point vortices are separated into $m = 5$ communities. In this problem, there are strong intra-community ties in the original adjacency matrix, which enables a clear decomposition of the group of vortices into communities. For each community, we can determine the vortical community centroid using \eq (\ref{eq:vort_centroid}). The resulting reduced network of these community centroids is shown in \fig \ref{f:pvd} (top right). The number of interactions have been reduced to $m(m-1)$. 

The adjacency matrix for the reduced community network is given by
\begin{equation}
   \tilde{A}_{kl}
   = \frac{ |\Gamma_l| } {2\pi \left|\boldsymbol{\xi}_k - \boldsymbol{\xi}_l\right|},\quad \Gamma_l = \sum_{j \in C_l} \gamma_j,
   \label{eq:reduced4}
\end{equation}
where the state vector $\boldsymbol{\xi}$ contains only the position information of the community centroids. The circulation $\Gamma$ and $\gamma$ of the vortices do not change in time as the flow is inviscid. Here, the edge weights for the reduced graph are quantified in the same manner as in the original network, but now for the community centroids. Note that the reduced network weights conserve the total circulation of the system, as all circulations within the community are summed. The reduced adjacency matrix is shown in \fig \ref{f:pvd} next to the reduced centroid network.  As the communities with strong intra-community ties are reduced to their representative centroids, the reduced adjacency matrix highlights the interactions between communities. We see a strong interaction between communities $1$ and $2$.  Community $3$ experiences weak influence from all other communities, except from community $4$, due to community $3$ being placed far from the rest of the communities.

The interaction-based dynamics of the discrete point vortex system is governed by the Biot-Savart law given by \eq (\ref{eq:BS}). This can be treated as a networked dynamical system given by \eq (\ref{eq:1}) with a time-varying network. As the position of the point vortices changes over time, the adjacency matrix can be updated accordingly.  Trajectories based on the full nonlinear dynamics of the full system with $n = 100$ vortices are shown in \fig \ref{f:pvd} (bottom left). The spatial location of the centroids are non-dimensionalized by the average radial distance of the centroids of the communities from the geometric center of the overall system at the initial time ($R_0$).  The time variable is non-dimensionalized by the total circulation $\Gamma_t = \sum_{j=1}^n\gamma_j$ and the average radial distance of the centroids of the clusters from the geometric center of the overall system at the initial time ($R_0$) as $t\Gamma_t/2\pi R_0^2$.  Shown in \fig \ref{f:pvd} (bottom left) are the vortical positions for the communities at representative times.  Also superposed in dashed lines are the trajectories of the community centroids ($\boldsymbol{\xi}_c$) to track their bulk motion.  Although, these centroids are defined according to \eq (\ref{eq:vort_centroid}), these are computed \textit{a posteriori} considering updated positions of all the vortices within the communities.    

Reducing the system to the community centroid-based representation, we can solve for the dynamics of the community centroids using \eq (\ref{eq:reduced3}).  For the community centroid dynamics of vortical flows, $\boldsymbol{f}(\boldsymbol{\xi}_k) = 0$ and the interaction function $\tilde{\boldsymbol{g}}$ takes the form of
\begin{equation}
\tilde{\boldsymbol{g}}(\boldsymbol{\xi}_k, \boldsymbol{\xi}_l) = \frac{\hat{\boldsymbol{k}} \times (\boldsymbol{\xi}_k- \boldsymbol{\xi}_l)}{|\boldsymbol{\xi}_k- \boldsymbol{\xi}_l|}.
\label{eq:reduced_dy}
\end{equation}
Note that $\boldsymbol{g}$ and $\tilde{\boldsymbol{g}}$ are of the same form. Using \eqs (\ref{eq:reduced4}) and (\ref{eq:reduced_dy}) in  \eq (\ref{eq:reduced3}), we can solve for the dynamics of the community centroids as shown in \fig \ref{f:pvd} (bottom right). Here again, time-varying reduced adjacency matrices are computed for the dynamics of the centroids.  Upon comparing the dynamics of the original system and the community-based reduced-order model, it can be seen that the community-based model accurately tracks the bulk motion of the communities. The errors in the centroid trajectories ($\|\boldsymbol{\xi}_c - \boldsymbol{\xi}\|/R_0$) determined by the community-based reduced-order model are shown in \fig \ref{f:pvd} (bottom center).  The error level is limited to $\mathcal{O}(10^{-2})$ for all communities over the course of the rotational motion.  The community-based model tracks the unsteady dynamics of all discrete vortex communities in an accurate manner.

The application of the network community-based reduced-order model to the analysis of point vortex dynamics has shown its effectiveness to accurately capture the inter-community interactions and the macroscopic dynamics of the overall system.  The fact that the present model can handle the nonlinear flow dynamics its shows potential to capture the essential physics for a variety of vortex dominated flows in a computationally tractable manner.  In the next example, we demonstrate that the present formulation can be used for unsteady wake flows that have a continuous flow field representation over the spatial domain.  Even with the added complexity, we show that this network-based model can serve as a basis to develop a nonlinear model to predict unsteady forces on the body.


\subsection{Reduced-order wake force modeling}\label{subsec:force_model}

Fluid flow past solid bodies leads to the formation of complex, unsteady, nonlinear flow modifications in the wake. The laminar or turbulent wake gives rise to the formation and interaction of vortical structures, causing unsteady aerodynamic forces on the body. Here, the objective is to understand the dynamics of these flows based on network-theoretic approaches to model and predict the forces on the body. Thus, we formulate reduced-order models for these observable (auxiliary) variables of the flow.  In particular, we consider the modeling of aerodynamic lift and drag forces ($C_L$ and $C_D$, respectively), using the properties of the community-based centroids, as outlined in Sec.~\ref{sec:approach}.

As an example, we choose to model as $\boldsymbol{y}$ in \eq (\ref{eq:y_reduced}) the lift and drag coefficients $(C_L,~C_D)$, defined as
\begin{equation}
    C_L \equiv \frac{L}{\frac{1}{2}\rho u_{\infty}^2 S} 
    \quad \text{and}\quad 
    C_D \equiv \frac{D}{\frac{1}{2}\rho u_{\infty}^2 S},
\end{equation}
respectively.  In the above non-dimensionalization, $L$ and $D$ are the lift and drag forces on the body, $\rho$ is the fluid density, $u_{\infty}$ is the freestream velocity, and $S$ is the characteristic body surface area. The formation and interaction of the vortical structures lead to action of impulsive forces on the body \cite{Saffman:92,newton2013n}.
Thus, following the definition of induced velocity in \eq (\ref{eq:BS_simple}), it is sensible to select the circulation $\Gamma$ and inverse of the position $1/|\boldsymbol{\xi}|$ of the vortex community centroids, as the basis functions in \eq (\ref{eq:y_regression}) to model $\dot{\boldsymbol{y}}$. Note that the state vector $\boldsymbol{\xi}$ contains only the position vector of the community centroids. The circulation of each community is measured over time and varies in time due to the vorticity profile diffusing across the community boundary. We also consider the equivalent circulation on the body $(\Gamma_{0})$  \cite{batchelor2000introduction,anderson2010fundamentals} to be a basis function, which is related to the unsteady lift force through the Kutta--Joukowski theorem.

A polynomial combination of the aforementioned basis functions, summed over the $m$ centroids, forms a library $\boldsymbol{\Theta}$ to solve for $\boldsymbol{B}$ in \eq (\ref{eq:y_regression}). The order of polynomial library depends on the complexity of the vortical community dynamics of the wake.  Stronger the nonlinearity in the wake caused by vortical communities, more complex the unsteady force on the body, and thus higher order combination of the basis functions is required to construct the reduced-order model. For all problems discussed in the current paper, the third-order polynomial library is found sufficient to capture both the vortical interactions as well as the multiple frequencies appearing in the forces. As such, we arrive at the reduced-order model for ${\boldsymbol{y}}$ as,
\begin{equation}
\dot{\boldsymbol{Y}} =  \boldsymbol{\Theta} \boldsymbol{B}= 
\begin{bmatrix}
   \boldsymbol{\Theta}_1 & \boldsymbol{\Theta}_2 & \boldsymbol{\Theta}_3 
\end{bmatrix}
\begin{bmatrix}
   \boldsymbol{B}_1 \\ \boldsymbol{B}_2 \\ \boldsymbol{B}_3
\end{bmatrix}
\label{eq:forcemodel}
\end{equation}
where
\begin{equation}
    \dot{\boldsymbol{Y}}^T 
    = \left [ \dot{\boldsymbol{y}}^T(t_1),~
    \dot{\boldsymbol{y}}^T(t_2) \dots
    \dot{\boldsymbol{y}}^T(t_q) \right]\\
\end{equation}
and
\begin{align}
    \boldsymbol{\Theta}_1 &= \sum_{i=0}^m \Big[ f_i^1,~ f_i^2 \Big] \\
    \boldsymbol{\Theta}_2 &= \sum_{i,j=0}^m \Big[ f_i^1 f_j^1,~ f_i^1 f_j^2, ~f_i^2 f_j^2 \Big] \\
    \boldsymbol{\Theta}_3 &= \sum_{i,j,k=0}^m \Big[ f_i^1 f_j^1 f_k^1,~  f_i^1 f_j^1 f_k^2,~ f_i^1 f_j^2 f_k^2, ~f_i^2 f_j^2 f_k^2 \Big]
\end{align}
given
\begin{equation}
    f^1_i = \frac{1}{|\boldsymbol{\xi}_i(t)|},
    \quad 
    f^2_i = \Gamma_i (t),
    \quad
    \Gamma_0 = - \frac{C_L (t)}{\rho u_\infty}.
\end{equation}
where $\Gamma_0$ is the equivalent circulation on the body due to lift force and $\boldsymbol{\xi}_0$ is the position of the body. In the current work, we set $\boldsymbol{\xi}_0 = 0$ and exclude terms that contain $1/|\boldsymbol{\xi}_0|$.

We construct matrices $\dot{\boldsymbol{Y}} \in \mathbb{R}^{q \times 2}$ and $\boldsymbol{\Theta} \in \mathbb{R}^{q \times 9}$ using $q$ discrete time-series data over a training period.  The coefficient matrix $\boldsymbol{B} \in \mathbb{R}^{9 \times 2}$ is determined in this study using the Moore--Penrose inverse.  One can alternatively use a sparse matrix solver such as LASSO \cite{tibshirani1996regression,hastie2009overview}.  For the problems we consider below, the flow fields are reduced to a small number of community centroids, and does not necessitate the use of sparse solvers.  For the  wake flow problems analyzed in the current work, a training period of three cycles is found to result in convergence of the coefficients.  To predict the forces, \eq (\ref{eq:forcemodel}) can then be time integrated with $\Gamma$ and $1/|\boldsymbol{\xi}|$ of the vortex community centroids provided from corresponding time-series snapshots. Below, we apply these reduced-order formulations to predict lift and drag forces due to fluid flow over a two-dimensional circular cylinder and an airfoil with a flap.

\subsubsection{Unsteady force generated by wake of a circular cylinder}\label{subsubsec:results_cyl}

\begin{figure}
  \centerline{\includegraphics[width=0.48\textwidth]{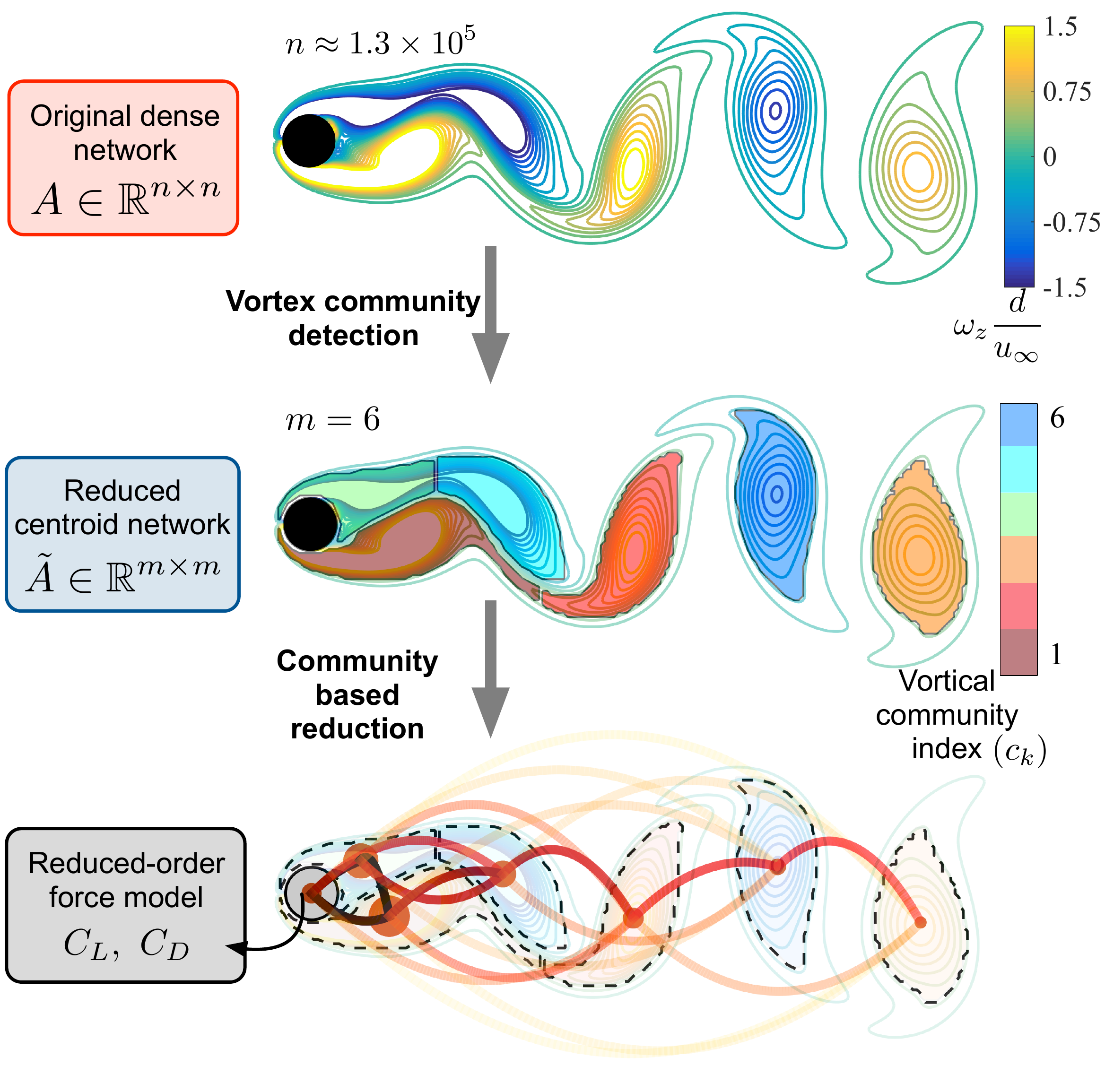}\quad}
  \vspace{-0.15in}
  \caption{Network community-based reduction of the vortices in the von K\'{a}rm\'{a}n vortex street and force modeling.}
\label{f:cyl_community}
\end{figure}

One of the fundamental phenomena in fluid dynamics is the formation of the von K\'{a}rm\'{a}n vortex street in bluff body wakes.  In particular, the wake of a circular cylinder, as depicted in \fig \ref{f:cyl_community}, serves as a canonical nonlinear problem in unsteady fluid dynamics. Even for a laminar regime, the shedding of the wake vortices leads to generation of large amplitude unsteady forces on the cylinder.  Although there have been extensive discussions on the mechanism of vortex shedding and its implication on unsteady forces \cite{rockwood2016detecting}, we focus here on simply feeding the flow field data to extract the vortical communities and use the centroid vector to develop a network-based force model.  That is, we rely on the network-based framework to decipher important vortical interactions and centroid quantities without depending on the full Navier--Stokes equations, as depicted in \fig \ref{f:cyl_community}.

The unsteady vortical flow field data to be used in this analysis is derived from two-dimensional direct numerical simulations (DNS), using the immersed boundary projection method \cite{Taira:JCP07,Colonius:CMAME08,Kajishima17}.  We consider incompressible flow over a circular cylinder at a diameter-based Reynolds number of $Re = 100$.  Extensive validation of this code and cylinder flow simulations have been performed and are reported in \cite{Munday:PF13}. Time resolved vorticity field snapshots are captured, and following the formulations in Sec.~\ref{subsec:biot_net}, the network of interactions among the vortical elements are defined for each snapshot using \eq (\ref{eq:adj_org}) \cite{Nair:JFM15, Taira:JFM16}. In this example, following the Eulerian description discussed in \fig \ref{f:flow_rep} to represent the continuous flow field, we take each and every Cartesian grid cell of size $\Delta x \times \Delta y$ to represent a network node with a circulation over the cell being denoted by $\gamma$. For the grid resolution chosen in the current study, the number of nodes and edges in the network scales to $\mathcal{O}(10^5)$ and $\mathcal{O}(10^{10})$ (following a vorticity threshold of $95\%$ of the maximum vorticity of original data, the number of nodes reduces to $\approx 4000$) respectively, as mentioned in \fig \ref{f:cyl_community}. Following the community detection procedure discussed in Sec.~\ref{subsec:fluid_community}, we identify different vortical communities in the wake as illustrated in \fig \ref{f:cyl_community} (middle), and further create a community-based network using the community centroids as shown in \fig \ref{f:cyl_community} (bottom). This reduces the number of nodes and edges to $6$ and $30$ (no self-loops), respectively. We also consider the cylinder itself as a node since the Kutta--Joukowski theorem relates lift to the circulation of the cylinder. 

Different vortical communities are shown in \fig \ref{f:cyl_community} (bottom) along with the centroids.  Here the nodes have varied sizes corresponding to their total network degree. The color and transparency of the lines connecting the nodes depict the strength of connection between them. The strong connection between adjacent communities in the wake is clearly depicted through this reduced dimensional network. Also, the high influence of the near-field vortical communities on the cylinder is evident.

\begin{figure}
  \centerline{\includegraphics[width=0.48\textwidth]{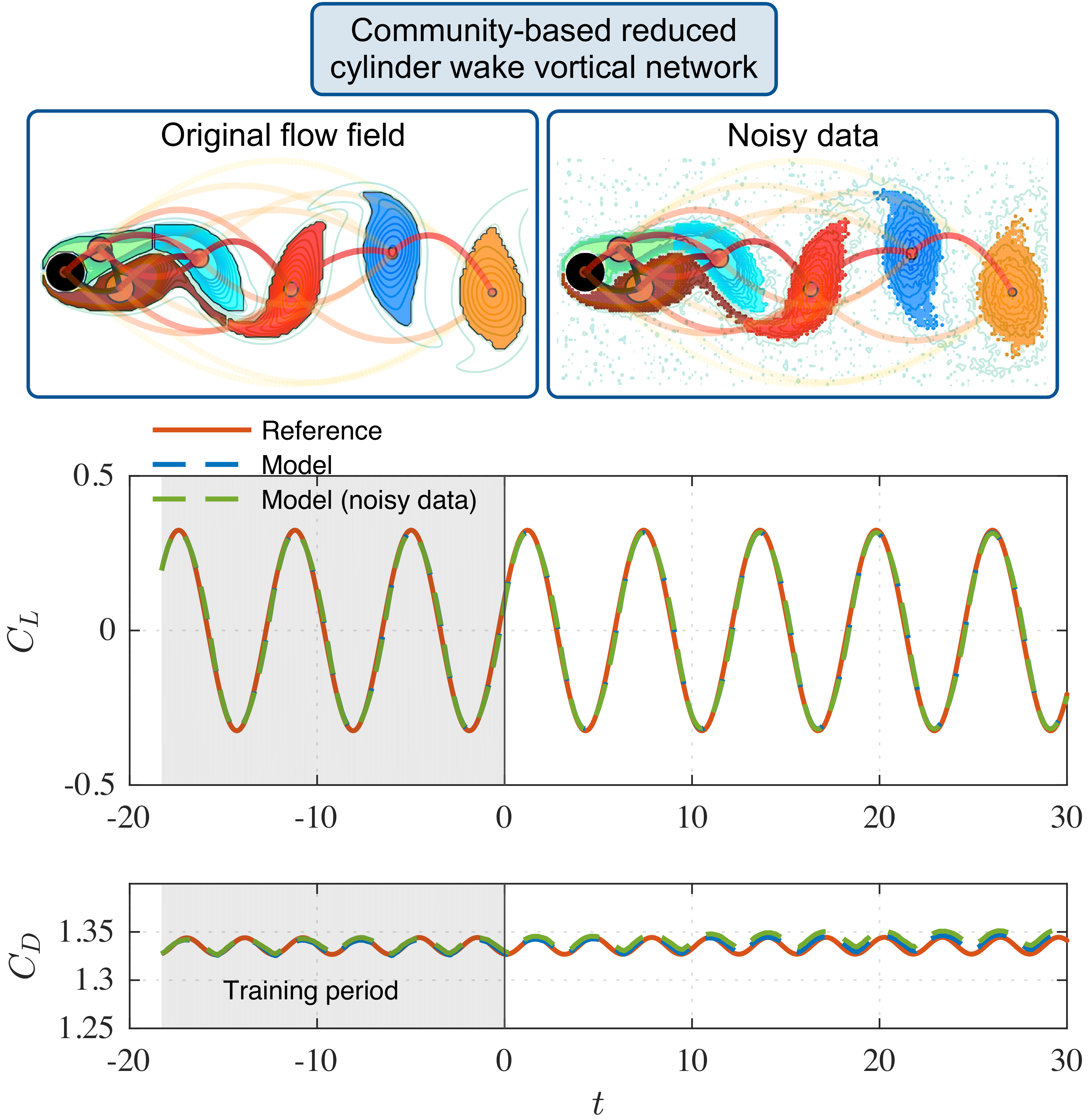}\quad}
\caption{Community-based reduced representation of a cylinder wake, and lift ($C_L$) and drag ($C_D$) forces on the cylinder predicted using the centroid-based reduced-order model. Resilience of the methodologies are tested using noisy flow field data.}
\label{f:results_cyl}
\end{figure}

To predict lift and drag on the cylinder, instantaneous training data of $\dot{\boldsymbol{y}}$ and the community-based vortical centroids from the flow fields are used to determine the nine coefficients in $\boldsymbol{B}$ in \eq (\ref{eq:forcemodel}). Due to the high influence of the near-field vortical communities and the time-periodic shedding of a positive vortex and a negative vortex, we only use the information of the first two community centroids ($m = 2$) to model the forces even though we show six vortical communities in the cylinder wake for visual clarity (\fig \ref{f:results_cyl} top). 
With coefficients $\boldsymbol{B}$ determined, \eq (\ref{eq:forcemodel}) is time integrated by providing time series of input data ($\boldsymbol{\xi}$ and $\Gamma$) to predict $C_L$ and $C_D$ on the cylinder. The predicted forces from this model are compared to the forces obtained from DNS in \fig \ref{f:results_cyl} (bottom), exhibiting agreement where predictions are on top of the reference forces.

The reduced-order force model, relying solely on the information of the community-based reduced vortical centroids, is able to accurately capture the effect of the vortex shedding phenomena on the forces. The results from \fig \ref{f:results_cyl} (bottom) show that the reduced-order force model accurately predicts the lift force with an error of $5.6\%$ for $L_2$ in time. The lift force is predicted over more than twice the training period while maintaining this accuracy, without any divergence later in time. The drag force, even with smaller level of fluctuations, is also accurately predicted over a considerable period of time with an error of $0.2\%$. Although the basis functions used depends on the vortical community formation (shedding) frequency, the nonlinear combinations of the basis functions chosen to arrive at \eq (\ref{eq:forcemodel}) is able to model the fluctuations in the drag force, which are twice the vortex shedding frequency for flow over a circular cylinder.

The resilience of the present model to perturbations or noise in the flow field (input data) is also examined.  Artificial noise with a signal-to-noise ratio of $SNR = 0.1$ is added to the original vorticity field data to simulate noise in experiments or turbulence in the flow. The community detection algorithm identifies the vortical communities from the noisy data as shown in \fig \ref{f:results_cyl} (top right). The network structures of the communities in the wake are still identified robustly even with the presence of noise. The forces are predicted using this noisy data as shown in \fig \ref{f:results_cyl} (bottom). The shown results also exhibit the resilience of the force-model towards noisy input data. The lift force prediction with the noisy input hardly deviates form that obtained using the original clean data with an error of $6.1\%$. The drag force is also predicted accurately with an error of $0.4\%$. The resilience of the model attributes to effects of both the vortical community detection and computation of the centroid-based properties acting as an integration filter, effectively removing the influence of external noise on the vorticity field.  As demonstrated above, the use of the reduced-order model allows for the unsteady vortical flow over a circular cylinder to be represented simply through the information distilled to the vortical centroids.

\subsubsection{Unsteady force on an airfoil with a Gurney flap}

\begin{figure}
  \centerline{\includegraphics[width=0.48\textwidth]{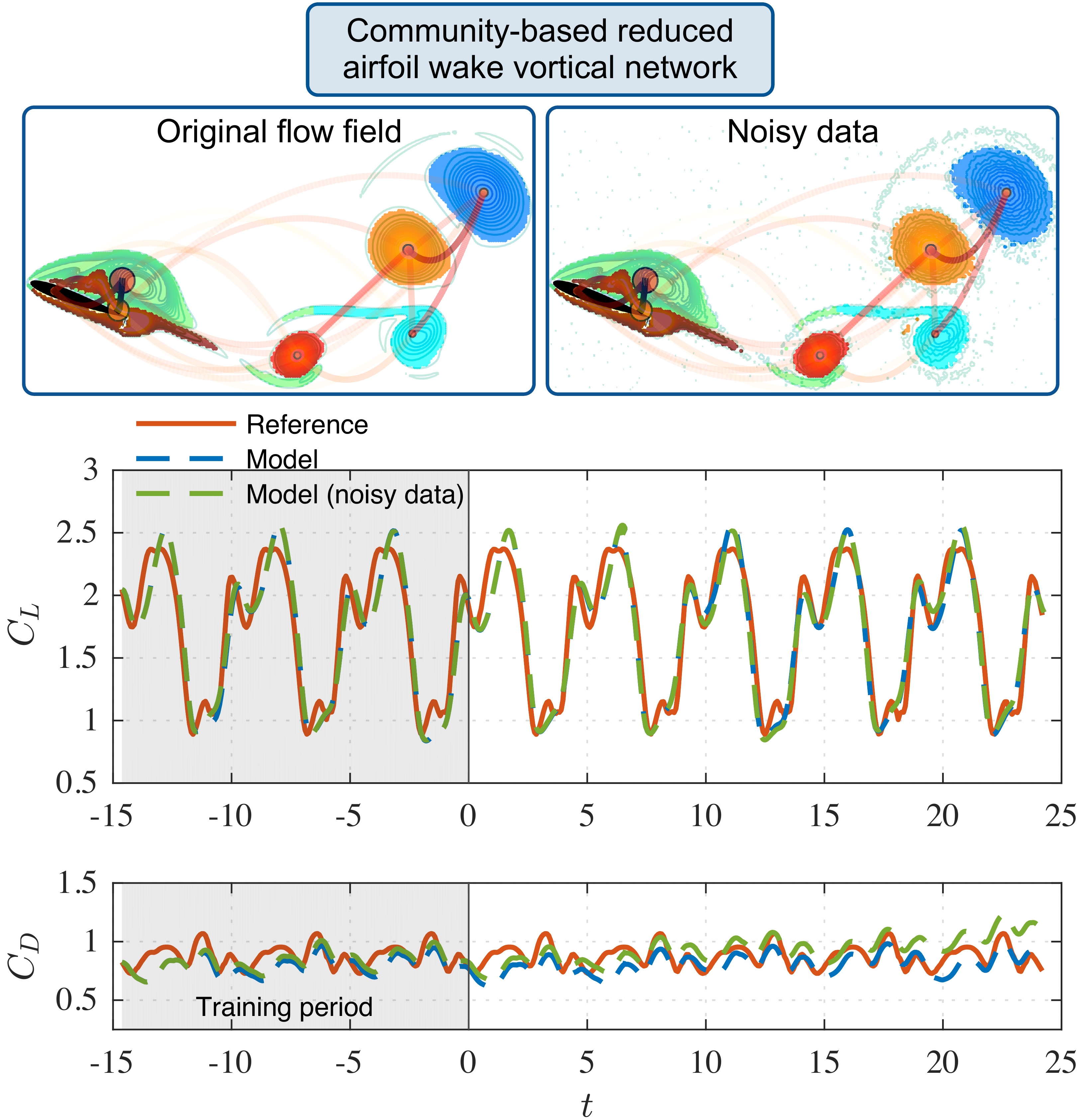}\quad}
\caption{Community-based reduced representation of a NACA 0012 airfoil wake classified under the 2P regime, and lift ($C_L$) and drag ($C_D$) forces on the airfoil predicted using the centroid-based reduced-order model. Resilience of the methodologies are tested using noisy flow field data. }\label{f:results_airfoil}
\end{figure}

Next, we extend the application of the present community-based formulation to a more complex flow. The intricate flow phenomena produced by the flow over an airfoil with a flow modification device, Gurney flap \cite{Liebeck:JOA78}, attached to the trailing edge is explored \cite{Meena:AIAAJ17}. Flow modification and control of airfoil wakes have been studied extensively for lift enhancement and drag reduction \cite{Joslin09}. Low-Reynolds number flows over airfoils with Gurney flaps at high angles of attack generate a variety of strongly nonlinear, unsteady vortical flows posing complex effects on lift and drag forces in both two- and three-dimensional settings \cite{Meena:AIAAJ17}. Such two-dimensional incompressible flows over NACA 0012 airfoils can produce a wake classified under the 2P regime \cite{Williamson:JFS88,Meena:AIAA17,Meena:AIAAJ17}, with periodic shedding of two pairs of positive and negative vortices as depicted in \fig \ref{f:results_airfoil} (top left). The vortex shedding attributes to a nonlinear aerodynamic force characteristics on the airfoil. We perform two-dimensional DNS of the flow over a NACA 0012 airfoil at an angle of attack of $20^{\circ}$ and flap height of $0.1c$ ($c$: chord length) with a chord-based Reynolds number of $Re = 1,000$ \cite{Meena:AIAAJ17} using the immersed boundary projection method mentioned in Sec.~\ref{subsubsec:results_cyl}.

Following the procedure used to analyze the cylinder wake, we capture time-resolved vorticity field snapshots of the flow, and form the full dense vortical network for the 2P wake flow. Through the community-based formulation, vortical communities are extracted from the full network and vortical centroids are defined as shown in \fig \ref{f:results_airfoil} (top).  Along with the considerable reduction in dimension attained, the complex nonlinear 2P wake with the two main and two secondary vortical structures are effectively captured by the network-theoretic techniques. The high influence of the two near-field vortical communities being formed on the airfoil is clearly depicted through the reduced network. The strong influence among the vortical communities in the far-field is also revealed through the community-based network.

The effect of this vortical flow on lift and drag forces are modeled following the procedure discussed for the cylinder problem. We use the position vector and circulation of the time periodically shedding four vortical centroids ($m=4$) to model the forces using \eq (\ref{eq:forcemodel}). We present the results of the lift and drag prediction compared to that from the DNS in \fig \ref{f:results_airfoil} (bottom). The complex 2P wake imposes highly fluctuating forces on the airfoil with multiple peaks. The reduced-order model predicts this highly nonlinear time-periodic lift curve accurately with an error of $8.8\%$ for $L_2$ in time. The multiple peaks with varying amplitudes imposed by the shedding of the primary and secondary vortical communities are effectively captured by the current community-based model. The drag force, with comparatively less magnitude and small-scale fluctuations, is also captured with an error of $12.7\%$.  These results highlight the importance of effectively identifying the vortical communities in the flow field. 

To test the resilience of the force model to noise, we follow the same procedure as in the cylinder problem by adding artificial noise to the flow field. Again, the voritcal communities are effectively captured by the community detection procedure as shown in \fig \ref{f:results_airfoil} (top right), preserving the network structure observed for the original noiseless flow field. A time series of this data is used to predict the forces and the results are depicted in \fig \ref{f:results_airfoil} (bottom). The lift force prediction based on the noisy flow field data remains accurate without noticeable difference from the prediction based on clean data with an error of $8.8\%$. The drag is also predicted well. The predicted drag curve deviates at later time, but the amplitude of deviation is small compared to the mean of the $C_D$ curve, and the error is $13.2\%$. The integrating effect of the community detection technique and the centroid-based calculations is further exemplified here. These results show the ability of the network community-based formulation to model forces on a body in a computationally reduced and robust manner.

\section{Conclusion}

We considered a network-community based reduce-order formulation to distill high-dimensional complex nonlinear interactions in unsteady vortical flows. The full vortical network representation (following either an Eulerian or Lagrangian description) is used to identify the communities or modular structures in the unsteady flows through the modularity maximization algorithm.  We then collapse the full vortical network through the centroid of each communities, considerably reducing the dimension of the fluid flow.  Furthermore, we extend the formulation to model and predict observable variables of the system, such as lift and drag on bodies, relying solely on the community-centroid properties.

The overall formulation is applied to canonical vortical flow problems for validation.  First, the current approach was applied to model the dynamics of a collection of discrete point vortices. Instead of tracking the dynamics of a large number of vortices, the overall dynamics of the discrete vortex communities is modeled using community-based networked dynamics.  Second, we have also modeled the lift and drag on bodies in wake flows using the community-based reduced-order formulation. Relying on the properties of the community centroids in the wake, models predicting lift and drag on the body are formulated through nonlinear regression analysis. These formulations are applied on the two-dimensional flow field data of flow over a circular cylinder and an airfoil with a Gurney flap. The community-based formulation is able to effectively identify the key vortical communities in both the wake flows. Also, the reduced-order force model is able to predict lift and drag forces on the cylinder as well as airfoil with considerable accuracy. The resilience of these formulations against perturbations or noise are also examined by adding artificial noise to the input data to simulate experimental noise or turbulence. The proposed formulation can serve as a basis to tackle highly chaotic and aperiodic flows. However, it is likely that techniques such as uncertainty quantification needs to be incorporated to assess the confidence of the model.

The demonstrations of the community-based reduced-order formulations, even though considered only for fluid flow problems in the current work, have implications to applicability to high-dimensional, nonlinear systems with intrinsic community-based coherent structures. The study alludes towards using the reduced-order methodology in effectively tracking the overall dynamics of such systems and modeling observable variables. Furthermore, these models can possibly be influenced by altering the interactions among the elements in the system. Coupling the formulations for the dynamics and observable variables may also motivate studies on control by altering the inter-community interactions, rather than examining the microscopic interactions within the communities.  The overall approach also comes with the advantages of formulating the models around the concept of understanding and representing the interactions among the elements through a data inspired network-theoretic methodology.

\begin{acknowledgments}

This work was supported by the National Science Foundation (Grant number: 1632003, Program Manager: Drs.~Dimitrios V.~Papavassiliou and Ronald D.~Joslin) and the Air Force Office of Scientific Research (Grant number: FA9550-16-1-0650, Program Manager: Dr.~Douglas R.~Smith).  We thank Prof.~Steven L.~Brunton for the stimulating discussions on reduced-order modeling during the course of this study.
\end{acknowledgments}

\bibliography{refs.bib}

\begin{thebibliography}{63}%
\makeatletter
\providecommand \@ifxundefined [1]{%
 \@ifx{#1\undefined}
}%
\providecommand \@ifnum [1]{%
 \ifnum #1\expandafter \@firstoftwo
 \else \expandafter \@secondoftwo
 \fi
}%
\providecommand \@ifx [1]{%
 \ifx #1\expandafter \@firstoftwo
 \else \expandafter \@secondoftwo
 \fi
}%
\providecommand \natexlab [1]{#1}%
\providecommand \enquote  [1]{``#1''}%
\providecommand \bibnamefont  [1]{#1}%
\providecommand \bibfnamefont [1]{#1}%
\providecommand \citenamefont [1]{#1}%
\providecommand \href@noop [0]{\@secondoftwo}%
\providecommand \href [0]{\begingroup \@sanitize@url \@href}%
\providecommand \@href[1]{\@@startlink{#1}\@@href}%
\providecommand \@@href[1]{\endgroup#1\@@endlink}%
\providecommand \@sanitize@url [0]{\catcode `\\12\catcode `\$12\catcode
  `\&12\catcode `\#12\catcode `\^12\catcode `\_12\catcode `\%12\relax}%
\providecommand \@@startlink[1]{}%
\providecommand \@@endlink[0]{}%
\providecommand \url  [0]{\begingroup\@sanitize@url \@url }%
\providecommand \@url [1]{\endgroup\@href {#1}{\urlprefix }}%
\providecommand \urlprefix  [0]{URL }%
\providecommand \Eprint [0]{\href }%
\providecommand \doibase [0]{http://dx.doi.org/}%
\providecommand \selectlanguage [0]{\@gobble}%
\providecommand \bibinfo  [0]{\@secondoftwo}%
\providecommand \bibfield  [0]{\@secondoftwo}%
\providecommand \translation [1]{[#1]}%
\providecommand \BibitemOpen [0]{}%
\providecommand \bibitemStop [0]{}%
\providecommand \bibitemNoStop [0]{.\EOS\space}%
\providecommand \EOS [0]{\spacefactor3000\relax}%
\providecommand \BibitemShut  [1]{\csname bibitem#1\endcsname}%
\let\auto@bib@innerbib\@empty
\bibitem [{\citenamefont {Bollob{\'a}s}(1998)}]{Bollobas98}%
  \BibitemOpen
  \bibfield  {author} {\bibinfo {author} {\bibfnamefont {B.}~\bibnamefont
  {Bollob{\'a}s}},\ }\href@noop {} {\emph {\bibinfo {title} {Modern graph
  theory}}}\ (\bibinfo  {publisher} {Springer},\ \bibinfo {year}
  {1998})\BibitemShut {NoStop}%
\bibitem [{\citenamefont {Newman}(2003)}]{Newman:SIAMReview03}%
  \BibitemOpen
  \bibfield  {author} {\bibinfo {author} {\bibfnamefont {M.~E.~J.}\
  \bibnamefont {Newman}},\ }\href@noop {} {\bibfield  {journal} {\bibinfo
  {journal} {SIAM Rev.}\ }\textbf {\bibinfo {volume} {45}},\ \bibinfo {pages}
  {167} (\bibinfo {year} {2003})}\BibitemShut {NoStop}%
\bibitem [{\citenamefont {Barab\'asi}(2016)}]{BarabasiNS16}%
  \BibitemOpen
  \bibfield  {author} {\bibinfo {author} {\bibfnamefont {A.-L.}\ \bibnamefont
  {Barab\'asi}},\ }\href@noop {} {\emph {\bibinfo {title} {Network Science}}}\
  (\bibinfo  {publisher} {Cambridge University Press},\ \bibinfo {year}
  {2016})\BibitemShut {NoStop}%
\bibitem [{\citenamefont {Zhu}\ \emph {et~al.}(2007)\citenamefont {Zhu},
  \citenamefont {Gerstein},\ and\ \citenamefont {Snyder}}]{zhu2007getting}%
  \BibitemOpen
  \bibfield  {author} {\bibinfo {author} {\bibfnamefont {X.}~\bibnamefont
  {Zhu}}, \bibinfo {author} {\bibfnamefont {M.}~\bibnamefont {Gerstein}}, \
  and\ \bibinfo {author} {\bibfnamefont {M.}~\bibnamefont {Snyder}},\
  }\href@noop {} {\bibfield  {journal} {\bibinfo  {journal} {Genes \&
  development}\ }\textbf {\bibinfo {volume} {21}},\ \bibinfo {pages} {1010}
  (\bibinfo {year} {2007})}\BibitemShut {NoStop}%
\bibitem [{\citenamefont {Sporns}(2011)}]{sporns2011human}%
  \BibitemOpen
  \bibfield  {author} {\bibinfo {author} {\bibfnamefont {O.}~\bibnamefont
  {Sporns}},\ }\href@noop {} {\bibfield  {journal} {\bibinfo  {journal} {Annals
  of the New York Academy of Sciences}\ }\textbf {\bibinfo {volume} {1224}},\
  \bibinfo {pages} {109} (\bibinfo {year} {2011})}\BibitemShut {NoStop}%
\bibitem [{\citenamefont {Otte}\ and\ \citenamefont
  {Rousseau}(2002)}]{otte2002social}%
  \BibitemOpen
  \bibfield  {author} {\bibinfo {author} {\bibfnamefont {E.}~\bibnamefont
  {Otte}}\ and\ \bibinfo {author} {\bibfnamefont {R.}~\bibnamefont
  {Rousseau}},\ }\href@noop {} {\bibfield  {journal} {\bibinfo  {journal}
  {Journal of Information Science}\ }\textbf {\bibinfo {volume} {28}},\
  \bibinfo {pages} {441} (\bibinfo {year} {2002})}\BibitemShut {NoStop}%
\bibitem [{\citenamefont {Albert}\ \emph {et~al.}(2004)\citenamefont {Albert},
  \citenamefont {Albert},\ and\ \citenamefont
  {Nakarado}}]{albert2004structural}%
  \BibitemOpen
  \bibfield  {author} {\bibinfo {author} {\bibfnamefont {R.}~\bibnamefont
  {Albert}}, \bibinfo {author} {\bibfnamefont {I.}~\bibnamefont {Albert}}, \
  and\ \bibinfo {author} {\bibfnamefont {G.~L.}\ \bibnamefont {Nakarado}},\
  }\href@noop {} {\bibfield  {journal} {\bibinfo  {journal} {Physical Review
  E}\ }\textbf {\bibinfo {volume} {69}},\ \bibinfo {pages} {025103} (\bibinfo
  {year} {2004})}\BibitemShut {NoStop}%
\bibitem [{\citenamefont {Boccaletti}\ \emph {et~al.}(2006)\citenamefont
  {Boccaletti}, \citenamefont {Latora}, \citenamefont {Moreno}, \citenamefont
  {Chavez},\ and\ \citenamefont {Hwang}}]{boccaletti2006complex}%
  \BibitemOpen
  \bibfield  {author} {\bibinfo {author} {\bibfnamefont {S.}~\bibnamefont
  {Boccaletti}}, \bibinfo {author} {\bibfnamefont {V.}~\bibnamefont {Latora}},
  \bibinfo {author} {\bibfnamefont {Y.}~\bibnamefont {Moreno}}, \bibinfo
  {author} {\bibfnamefont {M.}~\bibnamefont {Chavez}}, \ and\ \bibinfo {author}
  {\bibfnamefont {D.-U.}\ \bibnamefont {Hwang}},\ }\href@noop {} {\bibfield
  {journal} {\bibinfo  {journal} {Physics Reports}\ }\textbf {\bibinfo {volume}
  {424}},\ \bibinfo {pages} {175} (\bibinfo {year} {2006})}\BibitemShut
  {NoStop}%
\bibitem [{\citenamefont {Newman}\ \emph {et~al.}(2011)\citenamefont {Newman},
  \citenamefont {Barabasi},\ and\ \citenamefont {Watts}}]{newman2011structure}%
  \BibitemOpen
  \bibfield  {author} {\bibinfo {author} {\bibfnamefont {M.~E.~J.}\
  \bibnamefont {Newman}}, \bibinfo {author} {\bibfnamefont {A.-L.}\
  \bibnamefont {Barabasi}}, \ and\ \bibinfo {author} {\bibfnamefont {D.~J.}\
  \bibnamefont {Watts}},\ }\href@noop {} {\emph {\bibinfo {title} {The
  structure and dynamics of networks}}}\ (\bibinfo  {publisher} {Princeton
  University Press},\ \bibinfo {year} {2011})\BibitemShut {NoStop}%
\bibitem [{\citenamefont {Salath{\'e}}\ and\ \citenamefont
  {Jones}(2010)}]{Salathe:PLOSCB10}%
  \BibitemOpen
  \bibfield  {author} {\bibinfo {author} {\bibfnamefont {M.}~\bibnamefont
  {Salath{\'e}}}\ and\ \bibinfo {author} {\bibfnamefont {J.~H.}\ \bibnamefont
  {Jones}},\ }\href@noop {} {\bibfield  {journal} {\bibinfo  {journal} {PLOS
  Computational Biology}\ }\textbf {\bibinfo {volume} {6}},\ \bibinfo {pages}
  {e1000736} (\bibinfo {year} {2010})}\BibitemShut {NoStop}%
\bibitem [{\citenamefont {Albert}\ \emph {et~al.}(2000)\citenamefont {Albert},
  \citenamefont {Jeong},\ and\ \citenamefont {Barab\'asi}}]{Albert:Nature00}%
  \BibitemOpen
  \bibfield  {author} {\bibinfo {author} {\bibfnamefont {R.}~\bibnamefont
  {Albert}}, \bibinfo {author} {\bibfnamefont {H.}~\bibnamefont {Jeong}}, \
  and\ \bibinfo {author} {\bibfnamefont {A.-L.}\ \bibnamefont {Barab\'asi}},\
  }\href@noop {} {\bibfield  {journal} {\bibinfo  {journal} {Nature}\ }\textbf
  {\bibinfo {volume} {406}},\ \bibinfo {pages} {378} (\bibinfo {year}
  {2000})}\BibitemShut {NoStop}%
\bibitem [{\citenamefont {Dorogovtsev}\ and\ \citenamefont
  {Mendes}(2013)}]{dorogovtsev2013evolution}%
  \BibitemOpen
  \bibfield  {author} {\bibinfo {author} {\bibfnamefont {S.~N.}\ \bibnamefont
  {Dorogovtsev}}\ and\ \bibinfo {author} {\bibfnamefont {J.}~\bibnamefont
  {Mendes}},\ }\href@noop {} {\emph {\bibinfo {title} {Evolution of networks:
  From biological nets to the Internet and WWW}}}\ (\bibinfo  {publisher}
  {Oxford University Press},\ \bibinfo {year} {2013})\BibitemShut {NoStop}%
\bibitem [{\citenamefont {Nair}\ and\ \citenamefont
  {Taira}(2015)}]{Nair:JFM15}%
  \BibitemOpen
  \bibfield  {author} {\bibinfo {author} {\bibfnamefont {A.~G.}\ \bibnamefont
  {Nair}}\ and\ \bibinfo {author} {\bibfnamefont {K.}~\bibnamefont {Taira}},\
  }\href@noop {} {\bibfield  {journal} {\bibinfo  {journal} {Journal of Fluid
  Mechanics}\ }\textbf {\bibinfo {volume} {768}},\ \bibinfo {pages} {549}
  (\bibinfo {year} {2015})}\BibitemShut {NoStop}%
\bibitem [{\citenamefont {Taira}\ \emph {et~al.}(2016)\citenamefont {Taira},
  \citenamefont {Nair},\ and\ \citenamefont {Brunton}}]{Taira:JFM16}%
  \BibitemOpen
  \bibfield  {author} {\bibinfo {author} {\bibfnamefont {K.}~\bibnamefont
  {Taira}}, \bibinfo {author} {\bibfnamefont {A.~G.}\ \bibnamefont {Nair}}, \
  and\ \bibinfo {author} {\bibfnamefont {S.~L.}\ \bibnamefont {Brunton}},\
  }\href@noop {} {\bibfield  {journal} {\bibinfo  {journal} {Journal of Fluid
  Mechanics}\ }\textbf {\bibinfo {volume} {795}},\ \bibinfo {pages} {R2}
  (\bibinfo {year} {2016})}\BibitemShut {NoStop}%
\bibitem [{\citenamefont {Scarsoglio}\ \emph {et~al.}(2016)\citenamefont
  {Scarsoglio}, \citenamefont {Iacobello},\ and\ \citenamefont
  {Ridolfi}}]{scarsoglio2016complex}%
  \BibitemOpen
  \bibfield  {author} {\bibinfo {author} {\bibfnamefont {S.}~\bibnamefont
  {Scarsoglio}}, \bibinfo {author} {\bibfnamefont {G.}~\bibnamefont
  {Iacobello}}, \ and\ \bibinfo {author} {\bibfnamefont {L.}~\bibnamefont
  {Ridolfi}},\ }\href@noop {} {\bibfield  {journal} {\bibinfo  {journal}
  {International Journal of Bifurcation and Chaos}\ }\textbf {\bibinfo {volume}
  {26}},\ \bibinfo {pages} {1650223} (\bibinfo {year} {2016})}\BibitemShut
  {NoStop}%
\bibitem [{\citenamefont {Nair}\ \emph {et~al.}(2017)\citenamefont {Nair},
  \citenamefont {Brunton},\ and\ \citenamefont {Taira}}]{nair2017networked}%
  \BibitemOpen
  \bibfield  {author} {\bibinfo {author} {\bibfnamefont {A.~G.}\ \bibnamefont
  {Nair}}, \bibinfo {author} {\bibfnamefont {S.~L.}\ \bibnamefont {Brunton}}, \
  and\ \bibinfo {author} {\bibfnamefont {K.}~\bibnamefont {Taira}},\
  }\href@noop {} {\bibfield  {journal} {\bibinfo  {journal} {arXiv preprint
  arXiv:1706.06335}\ } (\bibinfo {year} {2017})}\BibitemShut {NoStop}%
\bibitem [{\citenamefont {Xie}\ \emph {et~al.}(2007)\citenamefont {Xie},
  \citenamefont {Li},\ and\ \citenamefont {Wang}}]{xie2007new}%
  \BibitemOpen
  \bibfield  {author} {\bibinfo {author} {\bibfnamefont {Z.}~\bibnamefont
  {Xie}}, \bibinfo {author} {\bibfnamefont {X.}~\bibnamefont {Li}}, \ and\
  \bibinfo {author} {\bibfnamefont {X.}~\bibnamefont {Wang}},\ }\href@noop {}
  {\bibfield  {journal} {\bibinfo  {journal} {Physica A: Statistical Mechanics
  and its Applications}\ }\textbf {\bibinfo {volume} {384}},\ \bibinfo {pages}
  {725} (\bibinfo {year} {2007})}\BibitemShut {NoStop}%
\bibitem [{\citenamefont {Musolesi}\ and\ \citenamefont
  {Mascolo}(2006)}]{musolesi2006community}%
  \BibitemOpen
  \bibfield  {author} {\bibinfo {author} {\bibfnamefont {M.}~\bibnamefont
  {Musolesi}}\ and\ \bibinfo {author} {\bibfnamefont {C.}~\bibnamefont
  {Mascolo}},\ }in\ \href@noop {} {\emph {\bibinfo {booktitle} {Proceedings of
  the 2nd international workshop on Multi-hop ad hoc networks: from theory to
  reality}}}\ (\bibinfo {organization} {ACM},\ \bibinfo {year} {2006})\ pp.\
  \bibinfo {pages} {31--38}\BibitemShut {NoStop}%
\bibitem [{\citenamefont {Mucha}\ \emph {et~al.}(2010)\citenamefont {Mucha},
  \citenamefont {Richardson}, \citenamefont {Macon}, \citenamefont {Porter},\
  and\ \citenamefont {Onnela}}]{mucha2010community}%
  \BibitemOpen
  \bibfield  {author} {\bibinfo {author} {\bibfnamefont {P.~J.}\ \bibnamefont
  {Mucha}}, \bibinfo {author} {\bibfnamefont {T.}~\bibnamefont {Richardson}},
  \bibinfo {author} {\bibfnamefont {K.}~\bibnamefont {Macon}}, \bibinfo
  {author} {\bibfnamefont {M.~A.}\ \bibnamefont {Porter}}, \ and\ \bibinfo
  {author} {\bibfnamefont {J.-P.}\ \bibnamefont {Onnela}},\ }\href@noop {}
  {\bibfield  {journal} {\bibinfo  {journal} {Science}\ }\textbf {\bibinfo
  {volume} {328}},\ \bibinfo {pages} {876} (\bibinfo {year}
  {2010})}\BibitemShut {NoStop}%
\bibitem [{\citenamefont {Newman}(2006{\natexlab{a}})}]{newman2006modularity}%
  \BibitemOpen
  \bibfield  {author} {\bibinfo {author} {\bibfnamefont {M.~E.~J.}\
  \bibnamefont {Newman}},\ }\href@noop {} {\bibfield  {journal} {\bibinfo
  {journal} {Proceedings of the National Academy of Sciences}\ }\textbf
  {\bibinfo {volume} {103}},\ \bibinfo {pages} {8577} (\bibinfo {year}
  {2006}{\natexlab{a}})}\BibitemShut {NoStop}%
\bibitem [{\citenamefont {Fortunato}(2010)}]{Fortunato:PR10}%
  \BibitemOpen
  \bibfield  {author} {\bibinfo {author} {\bibfnamefont {S.}~\bibnamefont
  {Fortunato}},\ }\href@noop {} {\bibfield  {journal} {\bibinfo  {journal}
  {Physics Reports}\ }\textbf {\bibinfo {volume} {486}},\ \bibinfo {pages} {75}
  (\bibinfo {year} {2010})}\BibitemShut {NoStop}%
\bibitem [{\citenamefont {Variano}\ \emph {et~al.}(2004)\citenamefont
  {Variano}, \citenamefont {McCoy},\ and\ \citenamefont
  {Lipson}}]{variano2004networks}%
  \BibitemOpen
  \bibfield  {author} {\bibinfo {author} {\bibfnamefont {E.~A.}\ \bibnamefont
  {Variano}}, \bibinfo {author} {\bibfnamefont {J.~H.}\ \bibnamefont {McCoy}},
  \ and\ \bibinfo {author} {\bibfnamefont {H.}~\bibnamefont {Lipson}},\
  }\href@noop {} {\bibfield  {journal} {\bibinfo  {journal} {Physical Review
  Letters}\ }\textbf {\bibinfo {volume} {92}},\ \bibinfo {pages} {188701}
  (\bibinfo {year} {2004})}\BibitemShut {NoStop}%
\bibitem [{\citenamefont {Hadjighasem}\ \emph {et~al.}(2016)\citenamefont
  {Hadjighasem}, \citenamefont {Karrasch}, \citenamefont {Teramoto},\ and\
  \citenamefont {Haller}}]{hadjighasem2016spectral}%
  \BibitemOpen
  \bibfield  {author} {\bibinfo {author} {\bibfnamefont {A.}~\bibnamefont
  {Hadjighasem}}, \bibinfo {author} {\bibfnamefont {D.}~\bibnamefont
  {Karrasch}}, \bibinfo {author} {\bibfnamefont {H.}~\bibnamefont {Teramoto}},
  \ and\ \bibinfo {author} {\bibfnamefont {G.}~\bibnamefont {Haller}},\
  }\href@noop {} {\bibfield  {journal} {\bibinfo  {journal} {Physical Review
  E}\ }\textbf {\bibinfo {volume} {93}},\ \bibinfo {pages} {063107} (\bibinfo
  {year} {2016})}\BibitemShut {NoStop}%
\bibitem [{\citenamefont {Schlueter-Kuck}\ and\ \citenamefont
  {Dabiri}(2017)}]{schlueter2017coherent}%
  \BibitemOpen
  \bibfield  {author} {\bibinfo {author} {\bibfnamefont {K.~L.}\ \bibnamefont
  {Schlueter-Kuck}}\ and\ \bibinfo {author} {\bibfnamefont {J.~O.}\
  \bibnamefont {Dabiri}},\ }\href@noop {} {\bibfield  {journal} {\bibinfo
  {journal} {Journal of Fluid Mechanics}\ }\textbf {\bibinfo {volume} {811}},\
  \bibinfo {pages} {468} (\bibinfo {year} {2017})}\BibitemShut {NoStop}%
\bibitem [{\citenamefont {Newman}(2010)}]{Newman:10}%
  \BibitemOpen
  \bibfield  {author} {\bibinfo {author} {\bibfnamefont {M.~E.~J.}\
  \bibnamefont {Newman}},\ }\href@noop {} {\emph {\bibinfo {title} {Networks:
  an introduction}}}\ (\bibinfo  {publisher} {Oxford Univ. Press},\ \bibinfo
  {year} {2010})\BibitemShut {NoStop}%
\bibitem [{\citenamefont {Kramer}\ \emph {et~al.}(2009)\citenamefont {Kramer},
  \citenamefont {Eden}, \citenamefont {Cash},\ and\ \citenamefont
  {Kolaczyk}}]{kramer2009network}%
  \BibitemOpen
  \bibfield  {author} {\bibinfo {author} {\bibfnamefont {M.~A.}\ \bibnamefont
  {Kramer}}, \bibinfo {author} {\bibfnamefont {U.~T.}\ \bibnamefont {Eden}},
  \bibinfo {author} {\bibfnamefont {S.~S.}\ \bibnamefont {Cash}}, \ and\
  \bibinfo {author} {\bibfnamefont {E.~D.}\ \bibnamefont {Kolaczyk}},\
  }\href@noop {} {\bibfield  {journal} {\bibinfo  {journal} {Physical Review
  E}\ }\textbf {\bibinfo {volume} {79}},\ \bibinfo {pages} {061916} (\bibinfo
  {year} {2009})}\BibitemShut {NoStop}%
\bibitem [{\citenamefont {Hecker}\ \emph {et~al.}(2009)\citenamefont {Hecker},
  \citenamefont {Lambeck}, \citenamefont {Toepfer}, \citenamefont
  {Van~Someren},\ and\ \citenamefont {Guthke}}]{hecker2009gene}%
  \BibitemOpen
  \bibfield  {author} {\bibinfo {author} {\bibfnamefont {M.}~\bibnamefont
  {Hecker}}, \bibinfo {author} {\bibfnamefont {S.}~\bibnamefont {Lambeck}},
  \bibinfo {author} {\bibfnamefont {S.}~\bibnamefont {Toepfer}}, \bibinfo
  {author} {\bibfnamefont {E.}~\bibnamefont {Van~Someren}}, \ and\ \bibinfo
  {author} {\bibfnamefont {R.}~\bibnamefont {Guthke}},\ }\href@noop {}
  {\bibfield  {journal} {\bibinfo  {journal} {Biosystems}\ }\textbf {\bibinfo
  {volume} {96}},\ \bibinfo {pages} {86} (\bibinfo {year} {2009})}\BibitemShut
  {NoStop}%
\bibitem [{\citenamefont {De~Smet}\ and\ \citenamefont
  {Marchal}(2010)}]{de2010advantages}%
  \BibitemOpen
  \bibfield  {author} {\bibinfo {author} {\bibfnamefont {R.}~\bibnamefont
  {De~Smet}}\ and\ \bibinfo {author} {\bibfnamefont {K.}~\bibnamefont
  {Marchal}},\ }\href@noop {} {\bibfield  {journal} {\bibinfo  {journal}
  {Nature Reviews Microbiology}\ }\textbf {\bibinfo {volume} {8}},\ \bibinfo
  {pages} {717} (\bibinfo {year} {2010})}\BibitemShut {NoStop}%
\bibitem [{\citenamefont {Nelles}(2013)}]{nelles2013nonlinear}%
  \BibitemOpen
  \bibfield  {author} {\bibinfo {author} {\bibfnamefont {O.}~\bibnamefont
  {Nelles}},\ }\href@noop {} {\emph {\bibinfo {title} {Nonlinear system
  identification: from classical approaches to neural networks and fuzzy
  models}}}\ (\bibinfo  {publisher} {Springer Science \& Business Media},\
  \bibinfo {year} {2013})\BibitemShut {NoStop}%
\bibitem [{\citenamefont {Newman}\ and\ \citenamefont
  {Girvan}(2004)}]{newman2004finding}%
  \BibitemOpen
  \bibfield  {author} {\bibinfo {author} {\bibfnamefont {M.~E.~J.}\
  \bibnamefont {Newman}}\ and\ \bibinfo {author} {\bibfnamefont
  {M.}~\bibnamefont {Girvan}},\ }\href@noop {} {\bibfield  {journal} {\bibinfo
  {journal} {Physical Review E}\ }\textbf {\bibinfo {volume} {69}},\ \bibinfo
  {pages} {026113} (\bibinfo {year} {2004})}\BibitemShut {NoStop}%
\bibitem [{\citenamefont {Newman}(2004{\natexlab{a}})}]{newman2004fast}%
  \BibitemOpen
  \bibfield  {author} {\bibinfo {author} {\bibfnamefont {M.~E.~J.}\
  \bibnamefont {Newman}},\ }\href@noop {} {\bibfield  {journal} {\bibinfo
  {journal} {Physical Review E}\ }\textbf {\bibinfo {volume} {69}},\ \bibinfo
  {pages} {066133} (\bibinfo {year} {2004}{\natexlab{a}})}\BibitemShut
  {NoStop}%
\bibitem [{\citenamefont {Leicht}\ and\ \citenamefont
  {Newman}(2008)}]{leicht2008community}%
  \BibitemOpen
  \bibfield  {author} {\bibinfo {author} {\bibfnamefont {E.~A.}\ \bibnamefont
  {Leicht}}\ and\ \bibinfo {author} {\bibfnamefont {M.~E.~J.}\ \bibnamefont
  {Newman}},\ }\href@noop {} {\bibfield  {journal} {\bibinfo  {journal}
  {Physical Review Letters}\ }\textbf {\bibinfo {volume} {100}},\ \bibinfo
  {pages} {118703} (\bibinfo {year} {2008})}\BibitemShut {NoStop}%
\bibitem [{\citenamefont {Newman}(2006{\natexlab{b}})}]{newman2006finding}%
  \BibitemOpen
  \bibfield  {author} {\bibinfo {author} {\bibfnamefont {M.~E.~J.}\
  \bibnamefont {Newman}},\ }\href@noop {} {\bibfield  {journal} {\bibinfo
  {journal} {Physical Review E}\ }\textbf {\bibinfo {volume} {74}},\ \bibinfo
  {pages} {036104} (\bibinfo {year} {2006}{\natexlab{b}})}\BibitemShut
  {NoStop}%
\bibitem [{\citenamefont {Blondel}\ \emph {et~al.}(2008)\citenamefont
  {Blondel}, \citenamefont {Guillaume}, \citenamefont {Lambiotte},\ and\
  \citenamefont {Lefebvre}}]{blondel2008fast}%
  \BibitemOpen
  \bibfield  {author} {\bibinfo {author} {\bibfnamefont {V.~D.}\ \bibnamefont
  {Blondel}}, \bibinfo {author} {\bibfnamefont {J.-L.}\ \bibnamefont
  {Guillaume}}, \bibinfo {author} {\bibfnamefont {R.}~\bibnamefont
  {Lambiotte}}, \ and\ \bibinfo {author} {\bibfnamefont {E.}~\bibnamefont
  {Lefebvre}},\ }\href@noop {} {\bibfield  {journal} {\bibinfo  {journal}
  {Journal of Statistical Mechanics: Theory and Experiment}\ }\textbf {\bibinfo
  {volume} {2008}},\ \bibinfo {pages} {P10008} (\bibinfo {year}
  {2008})}\BibitemShut {NoStop}%
\bibitem [{\citenamefont {Newman}(2013)}]{newman2013spectral}%
  \BibitemOpen
  \bibfield  {author} {\bibinfo {author} {\bibfnamefont {M.~E.~J.}\
  \bibnamefont {Newman}},\ }\href@noop {} {\bibfield  {journal} {\bibinfo
  {journal} {Physical Review E}\ }\textbf {\bibinfo {volume} {88}},\ \bibinfo
  {pages} {042822} (\bibinfo {year} {2013})}\BibitemShut {NoStop}%
\bibitem [{\citenamefont {Pons}\ and\ \citenamefont
  {Latapy}(2006)}]{pons2006computing}%
  \BibitemOpen
  \bibfield  {author} {\bibinfo {author} {\bibfnamefont {P.}~\bibnamefont
  {Pons}}\ and\ \bibinfo {author} {\bibfnamefont {M.}~\bibnamefont {Latapy}},\
  }\href@noop {} {\bibfield  {journal} {\bibinfo  {journal} {J. Graph
  Algorithms Appl.}\ }\textbf {\bibinfo {volume} {10}},\ \bibinfo {pages} {191}
  (\bibinfo {year} {2006})}\BibitemShut {NoStop}%
\bibitem [{\citenamefont {Zhang}\ and\ \citenamefont
  {Newman}(2015)}]{zhang2015multiway}%
  \BibitemOpen
  \bibfield  {author} {\bibinfo {author} {\bibfnamefont {X.}~\bibnamefont
  {Zhang}}\ and\ \bibinfo {author} {\bibfnamefont {M.~E.~J.}\ \bibnamefont
  {Newman}},\ }\href@noop {} {\bibfield  {journal} {\bibinfo  {journal}
  {Physical Review E}\ }\textbf {\bibinfo {volume} {92}},\ \bibinfo {pages}
  {052808} (\bibinfo {year} {2015})}\BibitemShut {NoStop}%
\bibitem [{\citenamefont {Brunton}\ \emph {et~al.}(2016)\citenamefont
  {Brunton}, \citenamefont {Proctor},\ and\ \citenamefont
  {Kutz}}]{Brunton:PNAS16}%
  \BibitemOpen
  \bibfield  {author} {\bibinfo {author} {\bibfnamefont {S.~L.}\ \bibnamefont
  {Brunton}}, \bibinfo {author} {\bibfnamefont {J.~L.}\ \bibnamefont
  {Proctor}}, \ and\ \bibinfo {author} {\bibfnamefont {J.~N.}\ \bibnamefont
  {Kutz}},\ }\href@noop {} {\bibfield  {journal} {\bibinfo  {journal}
  {Proceedings of the National Academy of Sciences}\ }\textbf {\bibinfo
  {volume} {113}},\ \bibinfo {pages} {3932} (\bibinfo {year}
  {2016})}\BibitemShut {NoStop}%
\bibitem [{\citenamefont {Bates}\ and\ \citenamefont
  {Watts}(1988)}]{bates1988nonlinear}%
  \BibitemOpen
  \bibfield  {author} {\bibinfo {author} {\bibfnamefont {D.~M.}\ \bibnamefont
  {Bates}}\ and\ \bibinfo {author} {\bibfnamefont {D.~G.}\ \bibnamefont
  {Watts}},\ }\href@noop {} {\emph {\bibinfo {title} {Nonlinear regression
  analysis and its applications}}},\ Vol.~\bibinfo {volume} {2}\ (\bibinfo
  {publisher} {Wiley Online Library},\ \bibinfo {year} {1988})\BibitemShut
  {NoStop}%
\bibitem [{\citenamefont {Kajishima}\ and\ \citenamefont
  {Taira}(2017)}]{Kajishima17}%
  \BibitemOpen
  \bibfield  {author} {\bibinfo {author} {\bibfnamefont {T.}~\bibnamefont
  {Kajishima}}\ and\ \bibinfo {author} {\bibfnamefont {K.}~\bibnamefont
  {Taira}},\ }\href@noop {} {\emph {\bibinfo {title} {Computational Fluid
  Dynamics: Incompressible Turbulent Flows}}}\ (\bibinfo  {publisher}
  {Springer},\ \bibinfo {year} {2017})\BibitemShut {NoStop}%
\bibitem [{\citenamefont {Saffman}(1992)}]{Saffman:92}%
  \BibitemOpen
  \bibfield  {author} {\bibinfo {author} {\bibfnamefont {P.~G.}\ \bibnamefont
  {Saffman}},\ }\href@noop {} {\emph {\bibinfo {title} {Vortex dynamics}}}\
  (\bibinfo  {publisher} {Cambridge university press},\ \bibinfo {year}
  {1992})\BibitemShut {NoStop}%
\bibitem [{\citenamefont {Traag}\ and\ \citenamefont
  {Bruggeman}(2009)}]{traag2009community}%
  \BibitemOpen
  \bibfield  {author} {\bibinfo {author} {\bibfnamefont {V.~A.}\ \bibnamefont
  {Traag}}\ and\ \bibinfo {author} {\bibfnamefont {J.}~\bibnamefont
  {Bruggeman}},\ }\href@noop {} {\bibfield  {journal} {\bibinfo  {journal}
  {Physical Review E}\ }\textbf {\bibinfo {volume} {80}},\ \bibinfo {pages}
  {036115} (\bibinfo {year} {2009})}\BibitemShut {NoStop}%
\bibitem [{\citenamefont {Esmailian}\ and\ \citenamefont
  {Jalili}(2015)}]{esmailian2015community}%
  \BibitemOpen
  \bibfield  {author} {\bibinfo {author} {\bibfnamefont {P.}~\bibnamefont
  {Esmailian}}\ and\ \bibinfo {author} {\bibfnamefont {M.}~\bibnamefont
  {Jalili}},\ }\href@noop {} {\bibfield  {journal} {\bibinfo  {journal}
  {Scientific Reports}\ }\textbf {\bibinfo {volume} {5}} (\bibinfo {year}
  {2015})}\BibitemShut {NoStop}%
\bibitem [{\citenamefont {Sugihara}\ \emph {et~al.}(2013)\citenamefont
  {Sugihara}, \citenamefont {Liu},\ and\ \citenamefont
  {Murata}}]{sugihara2013community}%
  \BibitemOpen
  \bibfield  {author} {\bibinfo {author} {\bibfnamefont {T.}~\bibnamefont
  {Sugihara}}, \bibinfo {author} {\bibfnamefont {X.}~\bibnamefont {Liu}}, \
  and\ \bibinfo {author} {\bibfnamefont {T.}~\bibnamefont {Murata}},\
  }\href@noop {} {\bibfield  {journal} {\bibinfo  {journal} {Transactions of
  the Japanese Society for Artificial Intelligence}\ }\textbf {\bibinfo
  {volume} {28}},\ \bibinfo {pages} {67} (\bibinfo {year} {2013})}\BibitemShut
  {NoStop}%
\bibitem [{\citenamefont {Newman}(2004{\natexlab{b}})}]{newman2004analysis}%
  \BibitemOpen
  \bibfield  {author} {\bibinfo {author} {\bibfnamefont {M.~E.~J.}\
  \bibnamefont {Newman}},\ }\href@noop {} {\bibfield  {journal} {\bibinfo
  {journal} {Physical Review E}\ }\textbf {\bibinfo {volume} {70}},\ \bibinfo
  {pages} {056131} (\bibinfo {year} {2004}{\natexlab{b}})}\BibitemShut
  {NoStop}%
\bibitem [{\citenamefont {Clauset}\ \emph {et~al.}(2004)\citenamefont
  {Clauset}, \citenamefont {Newman},\ and\ \citenamefont
  {Moore}}]{clauset2004finding}%
  \BibitemOpen
  \bibfield  {author} {\bibinfo {author} {\bibfnamefont {A.}~\bibnamefont
  {Clauset}}, \bibinfo {author} {\bibfnamefont {M.~E.~J.}\ \bibnamefont
  {Newman}}, \ and\ \bibinfo {author} {\bibfnamefont {C.}~\bibnamefont
  {Moore}},\ }\href@noop {} {\bibfield  {journal} {\bibinfo  {journal}
  {Physical Review E}\ }\textbf {\bibinfo {volume} {70}},\ \bibinfo {pages}
  {066111} (\bibinfo {year} {2004})}\BibitemShut {NoStop}%
\bibitem [{\citenamefont {Fortunato}\ and\ \citenamefont
  {Barth{\'e}lemy}(2007)}]{fortunato2007resolution}%
  \BibitemOpen
  \bibfield  {author} {\bibinfo {author} {\bibfnamefont {S.}~\bibnamefont
  {Fortunato}}\ and\ \bibinfo {author} {\bibfnamefont {M.}~\bibnamefont
  {Barth{\'e}lemy}},\ }\href@noop {} {\bibfield  {journal} {\bibinfo  {journal}
  {Proceedings of the National Academy of Sciences}\ }\textbf {\bibinfo
  {volume} {104}},\ \bibinfo {pages} {36} (\bibinfo {year} {2007})}\BibitemShut
  {NoStop}%
\bibitem [{\citenamefont {Reichardt}\ and\ \citenamefont
  {Bornholdt}(2004)}]{reichardt2004detecting}%
  \BibitemOpen
  \bibfield  {author} {\bibinfo {author} {\bibfnamefont {J.}~\bibnamefont
  {Reichardt}}\ and\ \bibinfo {author} {\bibfnamefont {S.}~\bibnamefont
  {Bornholdt}},\ }\href@noop {} {\bibfield  {journal} {\bibinfo  {journal}
  {Physical Review Letters}\ }\textbf {\bibinfo {volume} {93}},\ \bibinfo
  {pages} {218701} (\bibinfo {year} {2004})}\BibitemShut {NoStop}%
\bibitem [{\citenamefont {Reichardt}\ and\ \citenamefont
  {Bornholdt}(2006)}]{reichardt2006statistical}%
  \BibitemOpen
  \bibfield  {author} {\bibinfo {author} {\bibfnamefont {J.}~\bibnamefont
  {Reichardt}}\ and\ \bibinfo {author} {\bibfnamefont {S.}~\bibnamefont
  {Bornholdt}},\ }\href@noop {} {\bibfield  {journal} {\bibinfo  {journal}
  {Physical Review E}\ }\textbf {\bibinfo {volume} {74}},\ \bibinfo {pages}
  {016110} (\bibinfo {year} {2006})}\BibitemShut {NoStop}%
\bibitem [{\citenamefont {Newton}(2013)}]{newton2013n}%
  \BibitemOpen
  \bibfield  {author} {\bibinfo {author} {\bibfnamefont {P.~K.}\ \bibnamefont
  {Newton}},\ }\href@noop {} {\emph {\bibinfo {title} {The N-vortex problem:
  analytical techniques}}},\ Vol.\ \bibinfo {volume} {145}\ (\bibinfo
  {publisher} {Springer},\ \bibinfo {year} {2013})\BibitemShut {NoStop}%
\bibitem [{\citenamefont {Batchelor}(2000)}]{batchelor2000introduction}%
  \BibitemOpen
  \bibfield  {author} {\bibinfo {author} {\bibfnamefont {G.~K.}\ \bibnamefont
  {Batchelor}},\ }\href@noop {} {\emph {\bibinfo {title} {An introduction to
  fluid dynamics}}}\ (\bibinfo  {publisher} {Cambridge university press},\
  \bibinfo {year} {2000})\BibitemShut {NoStop}%
\bibitem [{\citenamefont {Anderson~Jr}(2010)}]{anderson2010fundamentals}%
  \BibitemOpen
  \bibfield  {author} {\bibinfo {author} {\bibfnamefont {J.~D.}\ \bibnamefont
  {Anderson~Jr}},\ }\href@noop {} {\emph {\bibinfo {title} {Fundamentals of
  aerodynamics}}}\ (\bibinfo  {publisher} {McGraw-Hill},\ \bibinfo {year}
  {2010})\BibitemShut {NoStop}%
\bibitem [{\citenamefont {Tibshirani}(1996)}]{tibshirani1996regression}%
  \BibitemOpen
  \bibfield  {author} {\bibinfo {author} {\bibfnamefont {R.}~\bibnamefont
  {Tibshirani}},\ }\href@noop {} {\bibfield  {journal} {\bibinfo  {journal}
  {Journal of the Royal Statistical Society. Series B (Methodological)}\ ,\
  \bibinfo {pages} {267}} (\bibinfo {year} {1996})}\BibitemShut {NoStop}%
\bibitem [{\citenamefont {Hastie}\ \emph {et~al.}(2009)\citenamefont {Hastie},
  \citenamefont {Tibshirani},\ and\ \citenamefont
  {Friedman}}]{hastie2009overview}%
  \BibitemOpen
  \bibfield  {author} {\bibinfo {author} {\bibfnamefont {T.}~\bibnamefont
  {Hastie}}, \bibinfo {author} {\bibfnamefont {R.}~\bibnamefont {Tibshirani}},
  \ and\ \bibinfo {author} {\bibfnamefont {J.}~\bibnamefont {Friedman}},\ }in\
  \href@noop {} {\emph {\bibinfo {booktitle} {The elements of statistical
  learning}}}\ (\bibinfo  {publisher} {Springer},\ \bibinfo {year} {2009})\
  pp.\ \bibinfo {pages} {9--41}\BibitemShut {NoStop}%
\bibitem [{\citenamefont {Rockwood}\ \emph {et~al.}(2016)\citenamefont
  {Rockwood}, \citenamefont {Taira},\ and\ \citenamefont
  {Green}}]{rockwood2016detecting}%
  \BibitemOpen
  \bibfield  {author} {\bibinfo {author} {\bibfnamefont {M.~P.}\ \bibnamefont
  {Rockwood}}, \bibinfo {author} {\bibfnamefont {K.}~\bibnamefont {Taira}}, \
  and\ \bibinfo {author} {\bibfnamefont {M.~A.}\ \bibnamefont {Green}},\
  }\href@noop {} {\bibfield  {journal} {\bibinfo  {journal} {AIAA Journal}\
  }\textbf {\bibinfo {volume} {55}},\ \bibinfo {pages} {15} (\bibinfo {year}
  {2016})}\BibitemShut {NoStop}%
\bibitem [{\citenamefont {Taira}\ and\ \citenamefont
  {Colonius}(2007)}]{Taira:JCP07}%
  \BibitemOpen
  \bibfield  {author} {\bibinfo {author} {\bibfnamefont {K.}~\bibnamefont
  {Taira}}\ and\ \bibinfo {author} {\bibfnamefont {T.}~\bibnamefont
  {Colonius}},\ }\href@noop {} {\bibfield  {journal} {\bibinfo  {journal}
  {Journal of Computational Physics}\ }\textbf {\bibinfo {volume} {225}},\
  \bibinfo {pages} {2118} (\bibinfo {year} {2007})}\BibitemShut {NoStop}%
\bibitem [{\citenamefont {Colonius}\ and\ \citenamefont
  {Taira}(2008)}]{Colonius:CMAME08}%
  \BibitemOpen
  \bibfield  {author} {\bibinfo {author} {\bibfnamefont {T.}~\bibnamefont
  {Colonius}}\ and\ \bibinfo {author} {\bibfnamefont {K.}~\bibnamefont
  {Taira}},\ }\href@noop {} {\bibfield  {journal} {\bibinfo  {journal}
  {Computer Methods in Applied Mechanics and Engineering}\ }\textbf {\bibinfo
  {volume} {197}},\ \bibinfo {pages} {2131} (\bibinfo {year}
  {2008})}\BibitemShut {NoStop}%
\bibitem [{\citenamefont {Munday}\ and\ \citenamefont
  {Taira}(2013)}]{Munday:PF13}%
  \BibitemOpen
  \bibfield  {author} {\bibinfo {author} {\bibfnamefont {P.~M.}\ \bibnamefont
  {Munday}}\ and\ \bibinfo {author} {\bibfnamefont {K.}~\bibnamefont {Taira}},\
  }\href@noop {} {\bibfield  {journal} {\bibinfo  {journal} {Physics of
  Fluids}\ }\textbf {\bibinfo {volume} {25}} (\bibinfo {year}
  {2013})}\BibitemShut {NoStop}%
\bibitem [{\citenamefont {Liebeck}(1978)}]{Liebeck:JOA78}%
  \BibitemOpen
  \bibfield  {author} {\bibinfo {author} {\bibfnamefont {R.~H.}\ \bibnamefont
  {Liebeck}},\ }\href@noop {} {\bibfield  {journal} {\bibinfo  {journal}
  {Journal of Aircraft}\ }\textbf {\bibinfo {volume} {15}},\ \bibinfo {pages}
  {547} (\bibinfo {year} {1978})}\BibitemShut {NoStop}%
\bibitem [{\citenamefont {Gopalakrishnan~Meena}\ \emph
  {et~al.}(2018)\citenamefont {Gopalakrishnan~Meena}, \citenamefont {Taira},\
  and\ \citenamefont {Asai}}]{Meena:AIAAJ17}%
  \BibitemOpen
  \bibfield  {author} {\bibinfo {author} {\bibfnamefont {M.}~\bibnamefont
  {Gopalakrishnan~Meena}}, \bibinfo {author} {\bibfnamefont {K.}~\bibnamefont
  {Taira}}, \ and\ \bibinfo {author} {\bibfnamefont {K.}~\bibnamefont {Asai}},\
  }\href@noop {} {\bibfield  {journal} {\bibinfo  {journal} {AIAA Journal}\
  }\textbf {\bibinfo {volume} {56}},\ \bibinfo {pages} {1348} (\bibinfo {year}
  {2018})}\BibitemShut {NoStop}%
\bibitem [{\citenamefont {Joslin}\ and\ \citenamefont
  {Miller}(2009)}]{Joslin09}%
  \BibitemOpen
  \bibinfo {editor} {\bibfnamefont {R.~D.}\ \bibnamefont {Joslin}}\ and\
  \bibinfo {editor} {\bibfnamefont {D.}~\bibnamefont {Miller}},\ eds.,\
  \href@noop {} {\emph {\bibinfo {title} {Fundamentals and applications of
  modern flow control}}},\ Progress in Astronautics and Aeronautics\ (\bibinfo
  {publisher} {{AIAA}},\ \bibinfo {year} {2009})\BibitemShut {NoStop}%
\bibitem [{\citenamefont {Williamson}\ and\ \citenamefont
  {Roshko}(1988)}]{Williamson:JFS88}%
  \BibitemOpen
  \bibfield  {author} {\bibinfo {author} {\bibfnamefont {C.~H.~K.}\
  \bibnamefont {Williamson}}\ and\ \bibinfo {author} {\bibfnamefont
  {A.}~\bibnamefont {Roshko}},\ }\href@noop {} {\bibfield  {journal} {\bibinfo
  {journal} {Journal of Fluids and Structures}\ }\textbf {\bibinfo {volume}
  {2}},\ \bibinfo {pages} {355} (\bibinfo {year} {1988})}\BibitemShut {NoStop}%
\bibitem [{\citenamefont {Gopalakrishnan~Meena}\ \emph
  {et~al.}(2017)\citenamefont {Gopalakrishnan~Meena}, \citenamefont {Taira},\
  and\ \citenamefont {Asai}}]{Meena:AIAA17}%
  \BibitemOpen
  \bibfield  {author} {\bibinfo {author} {\bibfnamefont {M.}~\bibnamefont
  {Gopalakrishnan~Meena}}, \bibinfo {author} {\bibfnamefont {K.}~\bibnamefont
  {Taira}}, \ and\ \bibinfo {author} {\bibfnamefont {K.}~\bibnamefont {Asai}},\
  }in\ \href@noop {} {\emph {\bibinfo {booktitle} {AIAA paper 2017-0543}}}\
  (\bibinfo {year} {2017})\BibitemShut {NoStop}%
\end{thebibliography}%

\end{document}